\documentclass[preprint]{aastex}

\usepackage{color}

\slugcomment{Accepted for publication in ApJ}

\shortauthors{Chung et al.}

\defcitealias{2006Sci...311.1129Y}{Paper~I}
\defcitealias{2011ApJ...743..149Y}{Paper~II} 
\defcitealias{2011ApJ...743..150Y}{Paper~III}
\defcitealias{2013ApJ...768..137Y}{Paper~IV} 
\defcitealias{2013ApJ...768..138K}{Paper~V}
\defcitealias{2010AJ....139.1566F}{F10}
\defcitealias{2001MNRAS.326..959C}{C01}

\begin{document}

\title{
{Nonlinear Color--Metallicity Relations of Globular Clusters. VI.\\
On Calcium II Triplet Based Metallicities of Globular Clusters in Early-type Galaxies}
}

\author{Chul Chung\altaffilmark{1}, Suk-Jin Yoon\altaffilmark{1,2}, Sang-Yoon Lee\altaffilmark{1,2}, and Young-Wook Lee\altaffilmark{1,2}} 
\affil{Department of Astronomy \& Center for Galaxy Evolution Research, Yonsei University, Seoul 120-749, Republic of Korea}

\email{sjyoon@galaxy.yonsei.ac.kr}

\altaffiltext{1}{
Center for Galaxy Evolution Research, Yonsei University, Seoul 120-749, {Republic of} Korea; sjyoon@galaxy.yonsei.ac.kr.}
\altaffiltext{2}{
Department of Astronomy, Yonsei University, Seoul 120-749, {Republic of} Korea; sjyoon@galaxy.yonsei.ac.kr.}

\begin{abstract}

The metallicity distribution function of globular clusters (GCs) in galaxies is a key to understanding galactic formation and evolution.
The calcium II triplet (CaT) index has recently become a popular metal abundance indicator thanks to its sensitivity to GC metallicity.
Here we revisit and assess the reliability of CaT as a metallicity indicator using our new stellar population synthesis simulations based on empirical, high-resolution fluxes.
The model shows that the CaT strength of old ($>$ 10 Gyr) GCs is proportional to ${\rm [Fe/H]}$ below $-0.5$.
In the modest metal-rich regime, however, CaT does not increase anymore with ${\rm [Fe/H]}$ due to the little contribution from coolest red giant stars to the CaT absorption.
The nonlinear nature of the color--$CaT$ relation is confirmed by the observations of GCs in nearby early-type galaxies.
This indicates that the CaT should be used carefully when deriving metallicities of metal-rich stellar populations.
Our results offer an explanation for the observed sharp difference between the color and $CaT$ distributions of GCs in the same galaxies.
We take this as an analogy to the view that metallicity--color and metallicity--Lick index nonlinearity of GCs is primarily responsible for their observed ``bimodal'' distributions of colors and absorption indices.

\end{abstract}

\keywords{globular clusters: general --- stars: abundances --- stars: evolution --- stars: horizontal-branch}

\section{INTRODUCTION}

Deriving metallicity of globular clusters (GCs) in early-type galaxies (ETGs) has always been an important issue since the discovery of bimodal color distributions of GC systems in ETGs.
Colors are direct proxies for metallicity of {old-aged} stellar populations, and thus color bimodality was interpreted as the bimodal metallicity distributions \citep[e.g.,][]{1991ARA&A..29..543H, 2004Natur.427...31W, 2006ARA&A..44..193B}.
However, after \citet{2006Sci...311.1129Y} claimed that the conversion between color and metallicity is nonlinear due mainly to the effect of hot horizontal-branch (HB) stars on the integrated colors, our subsequent studies (\citealt{2011ApJ...743..149Y}; \citealt{2013ApJ...768..137Y}) supported nonlinear color--metallicity relations (CMRs) based on the multiband photometry of GC color distributions of M87 and M84.
\citet[][]{2011ApJ...743..150Y} further derived unimodal metallicity distribution functions of external GC systems which are similar to those of halo field stars of their host galaxies.
In order to derive more accurate metallicity of GCs in ETGs, various spectroscopic observations have been performed \citep{2004ApJ...602..705P, 2005AJ....130.1315S, 2007AJ....134..391C, 2008MNRAS.386.1443B, 2010ApJ...708.1335W, 2011MNRAS.415.3393F, 2012ApJ...759..116P, 2012ApJ...757..184P, 2012MNRAS.427.2349C}, but, as confirmed in \citet[][]{2013ApJS..204....3C}, \citet[][M31 spheroidal component GCs]{2013ApJ...768..138K}, and Kim et al. (2015, Paper VII, NGC~5128 GCs), the index--metallicity relations (IMRs) are also not free from the effect of hot HB stars. 

Recently, calcium II triplet (CaT) has been getting attention in the determination of metallicity of GCs in ETGs \citep{2010AJ....139.1566F, 2011MNRAS.415.3393F, 2012MNRAS.426.1475U, 2013MNRAS.436.1172U, 2015MNRAS.446..369U, 2012ApJ...759L..33B} because, compared to its sensitivity to metallicity, it is relatively less sensitive to hot stars such as blue HB stars in simple stellar populations (SSPs).
CaT is one of the strongest absorption features in the near infrared (8498, 8542, and 8662~\AA) and very sensitive to metallicity and surface gravity of stars \citep{2001MNRAS.326..959C, 2001MNRAS.326..981C, 2002MNRAS.329..863C}.
Therefore, CaT has been widely used as a metallicity indicator of various stellar systems such as GCs and open clusters in the Milky Way (MW), and extragalactic GCs, as well as dwarf galaxies \citep[e.g.,][]{1988AJ.....96...92A, 2012A&A...544A.109C, 2010AJ....139.1566F, 2008MNRAS.383..183B}.
{However, as shown in the CaT models of \citet{2003MNRAS.340.1317V}, when the metallicity is high enough as ${\rm [M/H]}\geq -0.5$, the CaT is not sensitive to the metallicity anymore, rather sensitive to the slope of initial mass functions of stellar populations.
Thus, probing the validity of using the CaT as a direct metallicity indicator is one of the most important issues.} 
In this paper we present the CaT model for SSPs based on empirical SEDs, and attempt to test whether CaT is a reliable metallicity tracer or not. 
Section~2 gives the construction of our CaT SSP model.
In section~3, we present CaT definitions adopted in this paper and results of our CaT models. 
Section~4 compares our models with GCs in various ETGs.
Section~5 discusses the implications of the nonlinear $CaT$--metallicity relation on the CaT distributions of GCs.  

\section{CONSTRUCTION OF MODELS}

The CaT model presented in this paper has been constructed following the standard Yonsei evolutionary population synthesis (YEPS) model described in \citet{2013ApJS..204....3C, 2013ApJ...769L...3C}. 
The model is an extended version of our previous models to include the near infrared wavelengths.
The only difference is that instead of adopting BaSel 3.1 library \citep{2002A&A...381..524W}, we have adopted empirical, high-resolution SEDs.
The input parameters, such as the initial mass function of main-sequence stars, the mass dispersion of HB stars, the mass-loss efficiency along the giant branch stars, and the age of inner-halo GCs in the MW, are exactly the same as in Table~1 of \citet{2013ApJS..204....3C}.
 
The empirical spectra we have adopted in the model are the Cenarro \citep{2001MNRAS.326..959C, 2001MNRAS.326..981C, 2002MNRAS.329..863C} and the INDO-US spectral libraries \citep{2004ApJS..152..251V} to cover the near infrared wavelength regime.
The Cenarro library comprises spectra of 702 stars with the wavelength coverage running from 8348 to 9020~{\AA} with a resolution of 1.5~{\AA}.
The INDO-US library contains spectra of 1273 stars with the wavelength from 3460 to 9464~{\AA} with 1~{\AA} spectral resolution, but some of spectra have several wavelength intervals.
Of 702 and 1273 empirical spectra of the Cenarro and INDO-US, we have selected 602 and 694 spectra, respectively, that have well determined stellar parameters and fully cover the CaT absorption features without any spectral gaps.
We have transformed stellar parameters (i.e., effective temperature, gravity, and total metallicity) of the INDO-US library to those of the Cenarro library using the transform equations from INDO-US to MILES library of \citet[][]{2012MNRAS.424..157V}.
Since the Cenarro and MILES \citep{2006MNRAS.371..703S} share 369 stars in common and they show almost the same stellar parameters \citep{2007MNRAS.374..664C}, the stellar parameter transformation from INDO-US to MILES library is equivalent to the transformation from INDO-US to Cenarro library.

In order to incorporate empirical SEDs in our model, we need to {calibrate fluxes of the two libraries}. 
{We firstly design box-shaped filter ranging from 8610 to 8810~{\AA} because of the Cenarro spectra covering wavelengths from 8350 to 9150~{\AA}. 
Secondly, we derive magnitudes of empirical spectra within the filter. 
At the same time, using the same filter we derive the magnitude of BaSel 3.1 spectrum at a given stellar parameter. 
After that we compare every empirical spectrum with the theoretical spectrum and calibrate the continuum levels of empirical spectra in units of [$\rm erg$~$\rm s^{-1} cm^{-2}${\AA}$^{-1}$]. 
In this process we do not mix two libraries together, and we separately develop two models based on two libraries. 
Therefore, the effect of missing spectral range on broadband colors reported in \citet{2012MNRAS.424..157V} is negligible in our models because we deal with narrow wavelength region of CaT within a given spectral library.}
We apply these calibrated empirical SEDs to stars in our synthetic color--magnitude diagram described in \citet{2013ApJS..204....3C} and reproduce empirical SSP models for CaT.

{In this paper, we choose to use the total metallicity ${\rm [Z/H]}$ for the construction of our models.} 
{Therefore, t}he CaT models for the enhanced $\alpha$-elements are mimicked by increasing the total metallicity ${\rm [Z/H]}$.
The enhancement of $\alpha$-elements is assumed to be ${\rm [\alpha/Fe]}=0.3$ \citep{2010ApJ...708.1335W, 2012MNRAS.427.2349C}{, and all models are constructed using $\alpha$-elements enhanced Y$^2$-stellar evolutionary tracks \citep{2002ApJS..143..499K}. 
We assume that ${\rm [Z/H]}$ relates to ${\rm [Fe/H]}$ with the relations of ${\rm [Z/H]} = {\rm [Fe/H]} + 0.723 \times {\rm [\alpha/Fe]}$ for ${\rm [a/Fe]}=0.3$ \citep[see also,][]{2003MNRAS.339..897T, 2005ApJS..160..176L, 2015MNRAS.449.1177V}. 
The constant 0.723 is derived from the $\alpha$-elements enhancement patterns of Y$^2$-isochrones with ${\rm [\alpha/Fe]}=0.3$ which show 2 times enhancement in O, Ne, Na, Mg, Si, P, S, Cl, Ar, Ca, and Ti, and depletion in Al and Mn \citep[see Table~2 of][]{2013ApJS..204....3C}. 
We apply this relation to derive ${\rm [Z/H]}$ of all empirical stars, and construct CaT SSP models for ${\rm [\alpha/Fe]}=0.3$.}

\section{MEASUREMENT OF C\MakeLowercase{A}T AND MODEL RESULTS}

\begin{figure}
\includegraphics[angle=-90,scale=0.9]{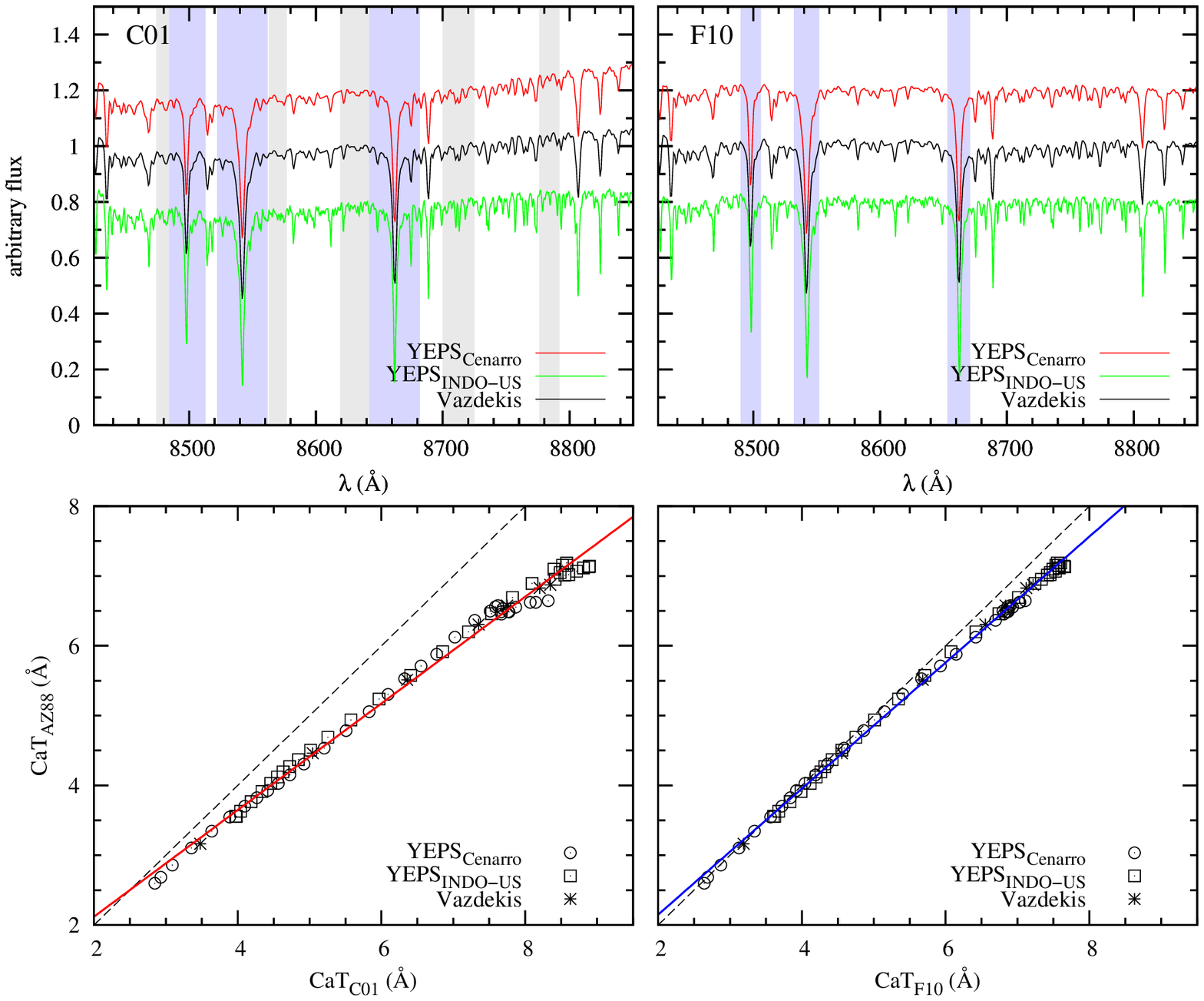}
\caption[]{
({\it Upper row}) Examples of model spectra before ({\it left}) and after fitting ({\it right}).  
Red and green lines are YEPS models based on the Cenarro and INDO-US libraries, respectively.
Selected models are $[{\rm Fe/H}]=0.0$ at 12~Gyr with ${\rm[\alpha/Fe]=0.3}$.
Black lines are the model of \citet{2003MNRAS.340.1317V} with similar metallicity and age ($[{\rm Fe/H}]=0.0$ and 12.5893~Gyr).
The highlighted blue and gray shades in the left panel are, respectively, the CaT index passbands and the continua of the Cenarro definition.
The blue shades in the right panel show the CaT index definition used in \citet{1988AJ.....96...92A} and \citet{2010AJ....139.1566F}.
({\it Bottom row}) Relations between CaT of \citet{1988AJ.....96...92A}, \citet{2001MNRAS.326..959C}, and \citet{2010AJ....139.1566F}.
The CaT absorptions are measured from our 13-Gyr models and 12.5893-Gyr models of \citet{2003MNRAS.340.1317V}, simultaneously.
Dashed lines are one to one correlation, and red and blue lines are the least-square fit of the data points.
}
\label{f1}
\end{figure}

There exist several index definitions of CaT when measuring its strength from observations \citep[][]{1984ApJ...283..457J, 1987A&A...186...49B, 1988AJ.....96...92A, 1989MNRAS.239..325D, 1991A&A...248..367Z, 1992AJ....103..711D, 1997PASP..109..883R, 2001MNRAS.326..959C, 2010AJ....139.1566F}.
Among these, we have chosen the definition of \citeauthor{2001MNRAS.326..959C}'s \citeyearpar[][C01]{2001MNRAS.326..959C} and \citeauthor{2010AJ....139.1566F}'s \citeyearpar[][F10]{2010AJ....139.1566F}.
The upper row of Figure~\ref{f1} shows the chosen CaT definitions along with our synthetic model spectra. 
The definition of \citetalias{2001MNRAS.326..959C} is designed for measuring the integrated CaT of galaxies and less affected by adjacent spectral features, such as Paschen and Fe I absorptions.
In order to compare our model with GC observations of \citetalias{2010AJ....139.1566F} and \citet{2012MNRAS.426.1475U}, we have adopted the method of \citetalias{2010AJ....139.1566F} which measures the value of $CaT$ from the normalized spectra whose continua are set to unity.
\citetalias{2010AJ....139.1566F} only measured the absorption features of central passbands ([8490.0--8506.0], [8532.0--8552.0], and [8653.0--8672.0]) of \citet{1988AJ.....96...92A} and did not consider continuum passbands of each index. 
Narrowing index passbands and setting continua as unity can also reduce the contamination of other adjacent absorption features \citep{2010AJ....139.1566F}.
We have followed the same way of the fitting and normalization of model spectra proposed in \citetalias{2010AJ....139.1566F} by using the IRAF {\small CONTINUUM} routine with Chebyshev polynomials of the order of eight.
{
Before we measure $CaT$ values, we apply Gaussian kernel smoothing to our model spectra based on Cenarro and INDO-US library in order to reduce the resolution of our model spectra to that of the Lick system. 
}
We will use the CaT index whose value was measured using the scheme of \citetalias{2010AJ....139.1566F} and \citetalias{2001MNRAS.326..959C} as ``$CaT_{\rm F10}$'' and ``$CaT_{\rm C01}$'', respectively, in what follows.

The bottom row of Figure~\ref{f1} provides the relations among $CaT_{\rm C01}$, $CaT_{\rm F10}$, and $CaT_{\rm AZ88}$ (the classical index definition of \citealt{1988AJ.....96...92A}).
The indices were measured from our 13-Gyr model SEDs and the 12.5893-Gyr SED of \citet{2003MNRAS.340.1317V}.
The measurement of $CaT_{\rm F10}$ only is taken from fitted spectra as shown in the top right panel.
For the conversion between our $CaT$ measurement and the other systems, we provide the best fitting relations in the following Equations~\ref{eq.1} (for the red line in the bottom row of Figure~\ref{f1}) and \ref{eq.2} (for the blue line in the bottom row of Figure~\ref{f1}).

\begin{equation}
{CaT_{\rm AZ88} = 0.7637 \times CaT_{\rm C01} + 0.5925}
\label{eq.1}
\end{equation}

\begin{equation}
{CaT_{\rm AZ88} = 0.9026 \times CaT_{\rm F10} + 0.3436}
\label{eq.2}
\end{equation}

The CaT models based on the Cenarro and INDO-US libraries are given in Figure~\ref{f2}.
{Our 3-, 5-, 10-, 12-, and 14-Gyr model predictions for CaT based on two empirical libraries are given in Table~\ref{tab.1}. 
Our simple stellar population models for CaT cover age ranging from 3 to 15~Gyr, metallicity from ${\rm [Fe/H]}=-2.5$ to $0.5$, and  scaled-solar abundance mixture and ${\rm [\alpha/Fe]}=0.3$.
The entire data are available at \url{http://web.yonsei.ac.kr/cosmic/data/YEPS.htm}.}

The $CaT$--${\rm [Fe/H]}$ relation for old age models ($\geq$~10~Gyr) shows an one-to-one correlation below ${\rm [Fe/H]=-0.5}$.
In the higher metallicity regime (${\rm [Fe/H]\geq-0.5}$), however, CaT does not trace ${\rm [Fe/H]}$ remaining constant around 8~{\AA} for $CaT_{\rm C01}$ and 7~{\AA} for $CaT_{\rm F10}$.
As a result, the overall shape of the relations is characterized by being nonlinear.
In order to find the physical cause for such nonlinearity, we first have checked whether hot HB stars affect the CaT strength as does for the optical colors \citepalias{2006Sci...311.1129Y, 2011ApJ...743..149Y, 2011ApJ...743..150Y, 2013ApJ...768..137Y} and spectral indices (\citealt{2013ApJS..204....3C}; \citetalias{2013ApJ...768..138K}; Paper~VII).
The age of models in the upper row of Figure~\ref{f2} is 12~Gyr, and the stellar populations at this age produce hot blue HB stars below ${\rm[Fe/H]} \sim -1.5$ \citep{2013ApJS..204....3C}. 
As shown in the Figure, the effect of hot HB stars ($>$~10,000~K) on $CaT$ is almost negligible for both $CaT_{\rm C01}$ and $CaT_{\rm F10}$.
This is caused by insensitivity of CaT to hot stars \citep{2002MNRAS.329..863C} and by the small flux contribution of hot HB stars in the near infrared regime. 
Our model shows that the flux contribution of HB stars to the total flux at 8700~{\AA} is only $\sim$5~\%, when the age and metallicity of SSP is 12~Gyr and ${\rm [Fe/H]}=-1.5$.
The effect of cool red HB stars on $CaT$ is a bit stronger than that of hot blue HB stars because the low gravity (${\rm log}\, g \leq 2.5$) and low temperature (${\rm T_{eff}} \sim 5000$~K) of those stars increase the strength of CaT \citep{2002MNRAS.329..863C}.
For instance, cool HB stars from metal-rich (${\rm [Fe/H]}=0.5$) stellar populations increase $CaT$ by up to 0.5~{\AA}.
This is however not enough to explain the cause of the nonlinear $CaT$--metallicity relation.

In the bottom row of Figure~\ref{f2} we have tested the effect of age on $CaT$. 
If the age of stellar population is greater than 10~Gyr, the model IMRs do not change significantly with age.
However, the CaT models of 3~Gyr show slightly different IMRs compared to the old age IMRs.
In the metal-poor regime, the younger age model shows enhanced $CaT$ due to the effect of hot turn-off stars that heavily contaminate Ca II by increasing the Paschen lines \citep{2002MNRAS.329..863C}.

\begin{figure}
\includegraphics[angle=-90,scale=0.85]{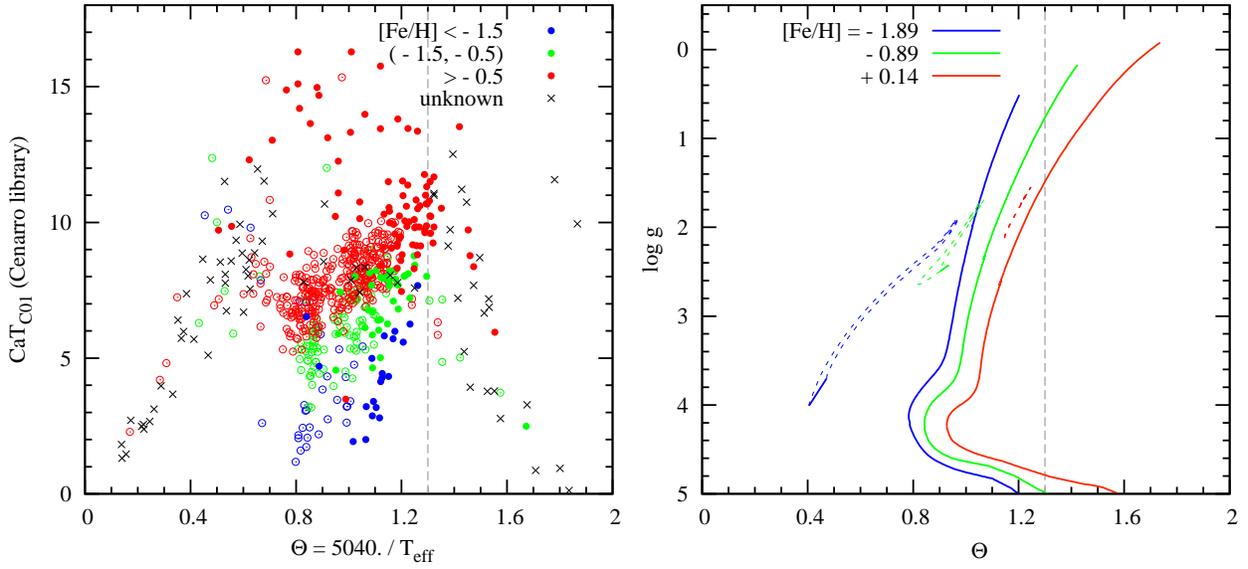}
\caption[]{
({\it Left}) The relation between temperature of stars and the strengths of CaT in the Cenarro library.
Metallicity of each star is denoted by colors.
Open and filled circles indicate dwarf (${\log g \geq 2.0}$) and giant (${< 2.0}$) stars, respectively.
Stellar library without sufficient information of metallicity or gravity are denoted as black crosses.
The vertical dashed grey line indicates $T_{\rm eff} = 3870$~K. 
({\it Right}) The Y$^2$-isochrones and HB tracks of 12~Gyr for different metallicities in the temperature and gravity plane.
{Colors of isochrones correspond to the metallicity bin in the left panel.}
The dashed grey line also indicates $T_{\rm eff} = 3870$~K.
}
\label{f3}
\end{figure}

Figure~\ref{f3} provides an explanation for nonlinearity of the CaT IMR.
In general, CaT is sensitive to metallicity and gravity of stars \citep{1997PASP..109..883R, 2009MNRAS.393..272W, 2015arXiv150700425V}.
These typical characteristics however are only for the stars whose temperature is between $T_{\rm eff} \sim$ 3870 and 7200~K.  
As shown in the left panel of Figure~\ref{f3}, when the temperature of stars is cooler than $T_{\rm eff} \sim$ 3870~K (grey dashed vertical lines in Figure~\ref{f3}), the strength of CaT gradually decreases with decreasing $T_{\rm eff}$ regardless of metallicity and gravity of stars.
The reason for this is that the formation of calcium II ionized absorption lines becomes less efficient in the lower temperature because calcium goes into a neutral state \citep{2002MNRAS.329..863C}.
As a result, if cool stars are dominant in a SSP, the strength of CaT can not be the indicator of metallicity and gravity of stellar populations.
The right panel of Figure~\ref{f3} shows that most of RGB stars whose ${\rm [Fe/H]}$ are greater than --0.89 are populated in the low temperature (${\rm T_{eff} < 3870}$~K) regime.
This explains why the strength of CaT for SSPs does not increase with increasing metallicity, converging to the same equivalent width (EW) around 8~{\AA} and 7~{\AA} for $CaT_{\rm C01}$ and $CaT_{\rm F10}$, respectively.  
We note that the similar $CaT$--metallicity relation is also found in the model of \citet[][in their Table~3]{2003MNRAS.340.1317V}.

\begin{figure}
\includegraphics[angle=-90,scale=0.85]{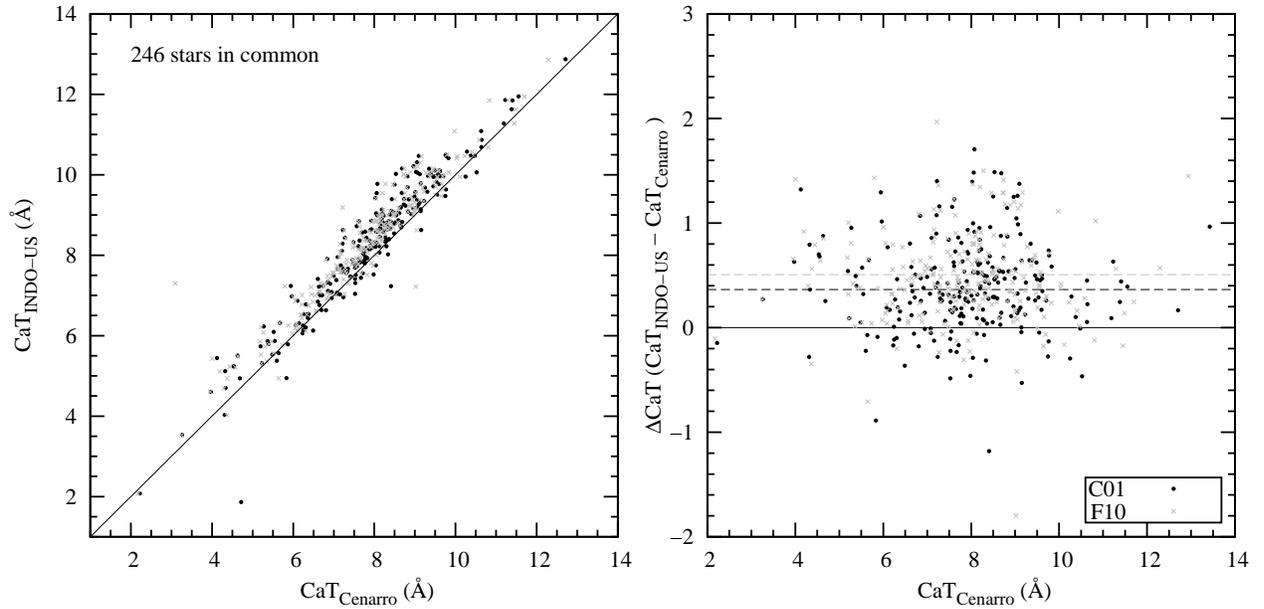}
\caption[]{
The correlation between $CaT_{\rm INDO-US}$ and $CaT_{\rm Cenarro}$ of stars in common (left) and the deviation of $CaT_{\rm INDO-US}$ from $CaT_{\rm Cenarro}$ (right).
Points are for 246 stars in common in Cenarro and INDO-US libraries.
Black and grey points are $CaT$ values which are measured in \citetalias{2001MNRAS.326..959C} and \citetalias{2010AJ....139.1566F} definitions.
Black lines are the one-to-one relations, and black and grey dashed line in the right panel is the arithmetic average of $CaT_{\rm INDO-US}$ deviation from $CaT_{\rm Cenarro}$.
}
\label{f4}
\end{figure}

{
In order to explain the offsets between Cenarro and INDO-US library, Figure~\ref{f4} compares the $CaT$ values of common stars in both libraries measured in \citetalias{2001MNRAS.326..959C} and \citetalias{2010AJ....139.1566F} definition.
We note that \citetalias{2010AJ....139.1566F} definition is relatively less affected by the flux calibration due to its flux normalization process before measuring $CaT$ value. 
Although we use two different definitions of CaT, the $CaT$ values measured in INDO-US library show, on average, 0.37 and 0.51~{\AA} enhanced $CaT$ in \citetalias{2001MNRAS.326..959C} and \citetalias{2010AJ....139.1566F} definitions, respectively. 
We speculate that this is due to the intrinsic difference {(such as inhomogeneous observations and different flux calibrations)} between empirical libraries and this is why the model based on INDO-US library shows $\sim$1 {\AA} offset from the model based on Cenarro library.
}

\begin{figure}
\includegraphics[angle=-90,scale=1.2]{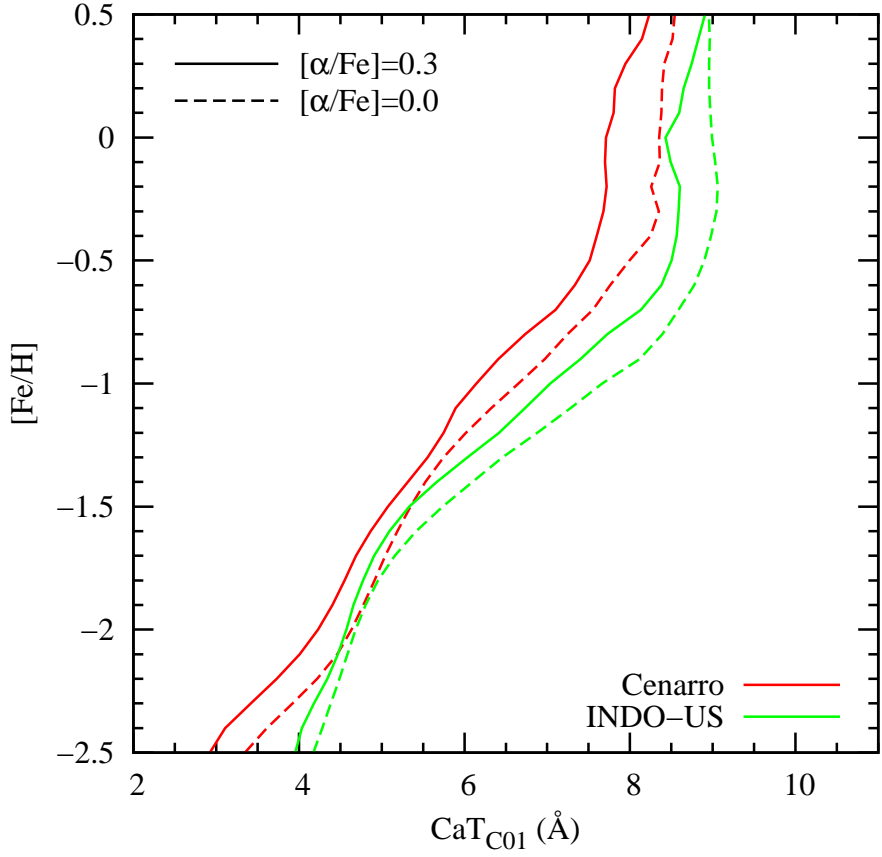}
\caption[]{
The result of CaT models based on the different assumption of $\alpha$-elements enhancement of empirical stars.
Solid and dashed lines are models for ${\rm [\alpha/Fe]}=0.3$ and 0.0 at the age of 12~Gyr.
Red and green lines are based on Cenarro and INDO-US library, respectively.
Regardless of the ${\rm [\alpha/Fe]}$, $CaT$ values do not increase with increasing metallicity in the metal-rich regime.
}
\label{f5}
\end{figure}

{
It is well established that there exists the well-known $\alpha$-element bias with respect to the metallicity of stars in the empirical libraries \citep[e.g.,][]{2013MNRAS.435..952S, 2015MNRAS.449.1177V}.
In Figure~\ref{f5}, we present ${\rm [\alpha/Fe]}=0.0$ model for CaT, which adopts scaled-solar stellar evolutionary tracks. 
In general, the scaled-solar model shows slightly enhanced $CaT$ strength compared to that of $\alpha$-elements enhanced models. 
If scaled-solar abundance for metal-rich populations (${\rm [Fe/H]}>-0.5$) is assumed, our models show, on average, 0.6~{\AA} enhanced CaT, but the saturation of CaT in the metal-rich regime still exists.
}

\section{COMPARISON WITH OBSERVATIONS}

\begin{figure}
\includegraphics[angle=-90,scale=0.9]{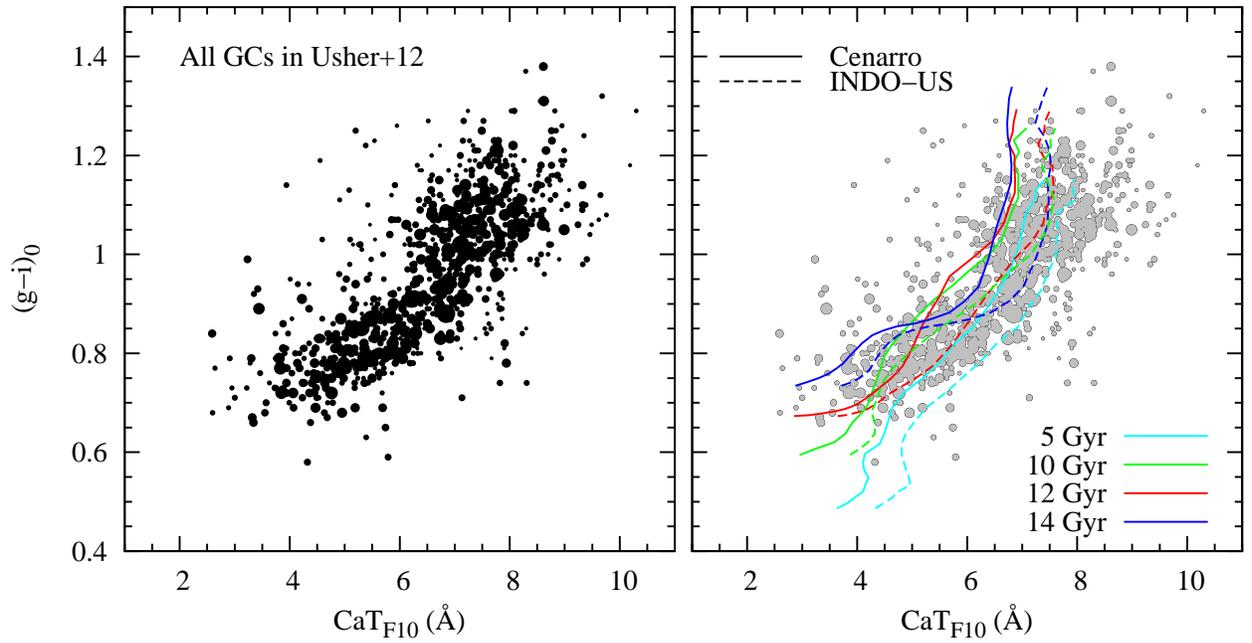}
\caption[]{
The $CaT$--$(g-i)_0$ relation of 901 GCs \citep{2012MNRAS.426.1475U} overlaid with our models.
The size of symbols is inversely proportional to the observational error.
Cyan, green, red, and blue lines are for 5, 10, 12, and 14~Gyr, respectively.
Solid and dashed lines are the models based on Cenarro and INDO-US libraries, respectively.
The $CaT$--$(g-i)_0$ relations are nonlinear both for observed and modeled GCs.

}
\label{f6}
\end{figure}

Figure~\ref{f6} shows the CaT versus $(g-i)_0$ diagram of 901 GCs in 11 ETGs in \citet{2012MNRAS.426.1475U} as part of the SAGES Legacy Unifying Globulars and Galaxies Survey\footnote{http://sluggs.swin.edu.au}. 
We have denoted $CaT$ of \citet{2012MNRAS.426.1475U} as $CaT_{\rm F10}$ as they followed the $CaT$ measurement of \citetalias{2010AJ....139.1566F}.
One can expect a linear correlation between $CaT_{\rm F10}$ and $(g-i)_0$ if both are direct metallicity tracers.
However, the observed relation is nonlinear steepening above $CaT_{\rm F10} \sim 6.5$~{\AA} and $(g-i) \sim 0.9$~mag.
This implies that at least one of two metallicity indicators is not directly tracing the metallicity of stellar populations.
Although the $(g-i)_0$--metallicity relation is inflected \citepalias{2006Sci...311.1129Y, 2011ApJ...743..149Y, 2013ApJ...768..137Y}, the $(g-i)_0$ color is just proportional to metallicity. 
However, as shown in Figure~\ref{f1}, our new IMR of CaT does not increase with increasing metallicity in the metal-rich regime.
This gives rise to the broken linear relation with different slopes below and above the point at $(g-i)_0 \sim 0.9$.

We overplot our models in the right panel of Figure~\ref{f6}.
The hot HB stars associated with metal-poor, old stellar populations make a ``wavy'' feature in the $CaT$--$(g-i)_0$ relation, but this effect disappears in the metal-rich regime.
Regardless of the choice of empirical libraries, our models show a converging feature around $CaT_{\rm F10} \sim$ 7~{\AA} because of the low formation efficiency of CaT for the most metal-rich stars.
As a consequence, the overall shapes of our old age models ($\geq 10$~Gyr) show reasonably good agreement with the observation.

\begin{figure}
\includegraphics[angle=0,scale=0.85]{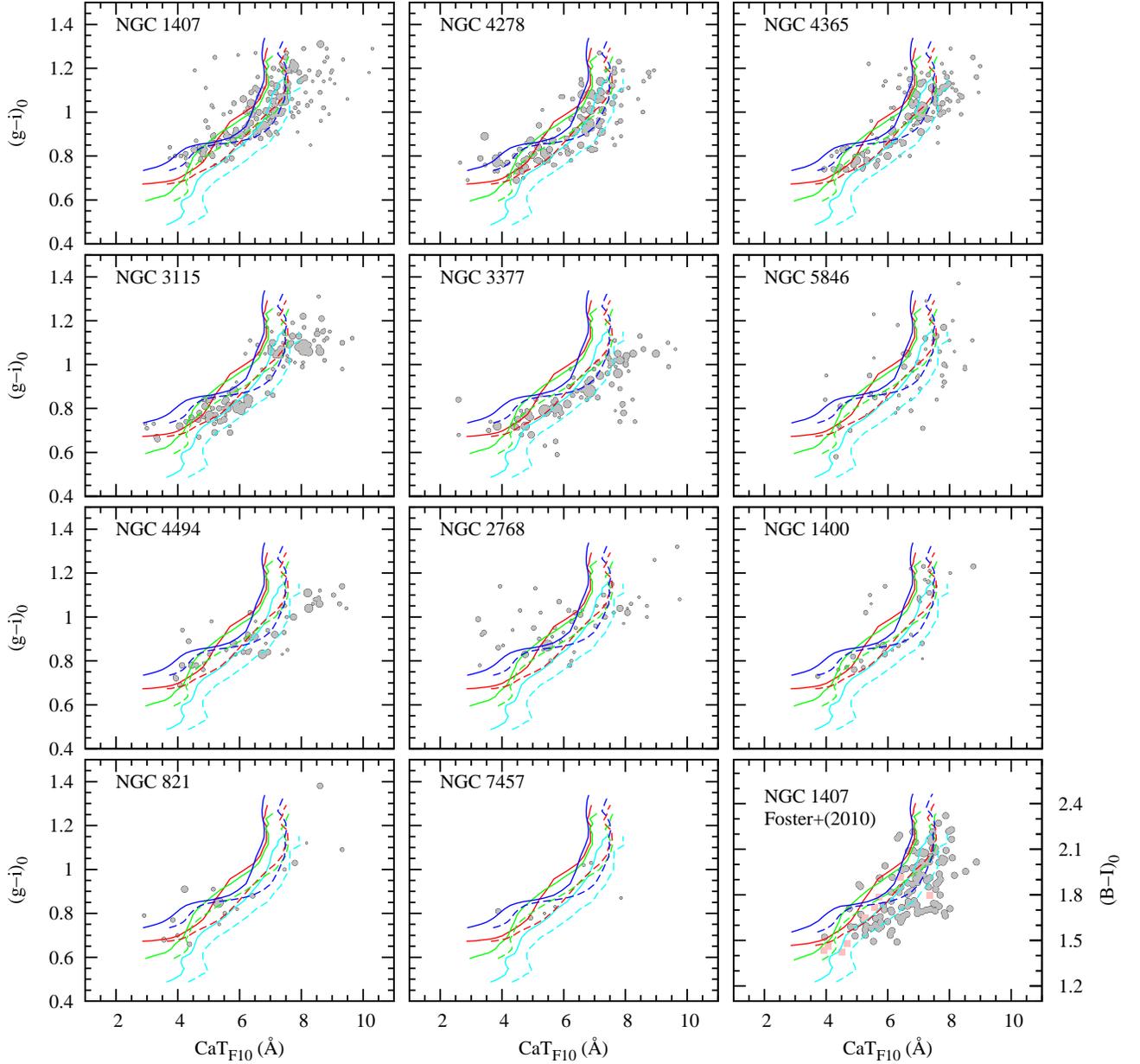}
\caption[]{
Same as Figure~\ref{f6} but for GCs in individual galaxies.
The GCs in rightmost bottom panel is from \citet{2010AJ....139.1566F}, and gray circles and red squares are GCs in NGC~1407 and the MW, respectively.
The size of symbols in this panel is inversely proportional to the observational error given in \citet{2010AJ....139.1566F}.
}
\label{f7}
\end{figure}

We compare our models with GCs of individual galaxies in Figure~\ref{f7}.
Our models show a good fit to most of GCs with small observational errors.
The GCs in NGC~1407, NGC~4278, and NGC~4365 well follow old age models.
The GCs of NGC~3115 show a clear separation between metal-poor and rich GCs, yet GCs still lie along with our models.
As suggested by \citet{2014A&A...564L...3C}, several metal-rich GCs in NGC~3115 prefer the younger age model (cyan lines), and this explains the presence of a group of metal-rich GCs with small offsets from models.
In the same vein, the metal-rich GCs in NGC~4494 can be also explained by younger age stellar populations \citep{2011MNRAS.415.3393F}.
The GC system in NGC~3377 does not have enough metal-rich GCs but in the metal-poor regime it agrees our models well.
The GCs in NGC~5846, NGC~2768, NGC~1400, NGC~821, and NGC~7457 are located along with our old age model predictions, although they have the small GC numbers. 
The CaT of GCs in NGC~1407 and the MW by \citetalias{2010AJ....139.1566F} is plotted in the rightmost bottom panel.
They also show good agreement with our models in the ${(B-I)}_0$ and CaT plane.

\section{DISCUSSION}

We have presented the CaT model for SSPs based on the high-resolution empirical SEDs.
We have demonstrated that the well-known metallicity indicator CaT loses its sensitivity to metallicity in the metal-rich regime, showing the highly nonlinear $CaT$--metallicity relation. 
Our model shows a good agreement with the observed GCs in the MW and ETGs.
If the underlying IMR of GC CaT is linear, this agreement between our model and the observation is hard to be explained.
Most studies however have adopted the linear one-to-one correlation for the conversion of $CaT$ to metallicity \citep[e.g.,][]{2010AJ....139.1566F, 2011MNRAS.415.3393F, 2012ApJ...759L..33B, 2012MNRAS.426.1475U, 2013MNRAS.436.1172U}.

An implication of the nonlinear IMR of CaT lies in the CaT distribution of GCs in ETGs.
As \citetalias{2006Sci...311.1129Y}, \citetalias{2011ApJ...743..149Y}, and \citetalias{2013ApJ...768..137Y} showed the case of optical colors, the nonlinear $CaT$--metallicity relation is most likely to reproduce the observed CaT distributions of GCs in ETGs from a single-peaked, broad metallicity distribution function (MDF).
Figures~\ref{f8} and \ref{f9} show the simulations for GC samples of \citet{2012MNRAS.426.1475U} and \citetalias{2010AJ....139.1566F} using our $CaT$--metallicity and color--metallicity relations.
We simulate the color distributions of GCs and find the best Gaussian MDFs for the observed distributions.
Then we simply apply those Gaussian MDFs to our empirical $CaT$--metallicity relations reproducing the CaT distributions.
The number of model GCs for the Gaussian MDF is $10^6$, and the observational uncertainties are taken into account in the projected $CaT$, ${(g-i)_0}$, and ${(B-I)}_0$ distributions.
Based on the observed GCs and $10^4$ randomly selected model GCs, we perform the Gaussian Mixture Modeling (GMM) analysis \citep{2010ApJ...718.1266M}.
Table~\ref{tab.2} summarizes the result for Figures~\ref{f8} and \ref{f9}.

{In the top panel of Figure~\ref{f8}, we perform the projection simulation for all 901 GCs of \citet[][]{2012MNRAS.426.1475U}.}
{The shapes of the observed $(g-i)_0$ distributions of 901 GCs are well reproduced by our model, when the GC MDF is assumed to be {${\rm \left<[Fe/H]\right>=-0.80}$} with $\sigma_{\rm [Fe/H]}=0.55$ and the age of the SSP is some 13~Gyr.}
The GMM analysis for the $(g-i)_0$ observation gives the peaks at 0.812 and 1.051~mag with the number fraction of 39.67 and 60.33~\%, respectively. 
The simulated $(g-i)_0$ distribution shows similar two peaks at 0.802 and 1.071~mag with the fraction of 31.10 and 68.90~\%, respectively.
The observed CaT distribution shows the metal-rich peak at 7.293~{\AA} and the metal-poor bump at 5.087~{\AA} with the fraction of 34.28 and 65.72~\%, respectively. 
{The same MDF produces a bimodal CaT histogram with at {5.224 and 6.704~{\AA}} with} the fraction of {53.25 and 46.75~\%}, respectively, based on the Cenarro IMR.
The prominent metal-rich peaks can be explained by the nonlinear IMR of CaT that makes metal-rich GCs pile up at $CaT \sim 6.7$~{\AA}.
Slight offsets between the observed and simulated CaT distribution may be attributed to the incompleteness of the population synthesis model in terms of stellar evolutionary tracks and/or empirical SEDs.

{In the second row of Figure~\ref{f8}, we present histograms for 779 GCs without NGC~3115 GCs.
NGC~3115 are well known for hosting two metallicity groups of GCs \citep{2011ApJ...736L..26A, 2012ApJ...759L..33B, 2014A&A...564L...3C, 2014AJ....148...32J}.
To see the pure effect of the nonlinear $CaT$--metallicity relation on the CaT distribution, we take out 122 GCs in NGC~3115 which may potentially mislead the analysis of the morphology of GC distributions.
A slightly smaller MDF dispersion with the same mean metallicity ($\sigma_{\rm [Fe/H]} = 0.50$ and $\left< {\rm [Fe/H]} \right> = -0.80$) reproduces both the $(g-i)_0$ and $CaT_{\rm F10}$ histograms, simultaneously.
A GMM bimodal fitting for $(g-i)_0$ yields 0.817 and 1.041 mag with 37.01 and 62.99~\% number fraction, respectively, for the observed histogram, and 0.806 and 1.064 mag with 30.38 and 69.62~\% number fraction, respectively, for the simulated one.
The exclusion of NGC~3115 GCs from all GC sample makes the observed metal-poor peak in the CaT histogram of GCs slightly weaker.
Our model reproduces this observation well. 
A GMM bimodal fitting for $CaT_{\rm F10}$ gives 4.686 and 7.090~{\AA} peaks with 27.13 and 72.87~\% number fraction, respectively, for the observed histogram, and 6.076 and 7.877~{\AA} peaks with 38.07 and 61.93~\% number fraction, respectively, for the simulated one based on the INDO-US IMR.
}

We also perform the projection simulations for GC systems of individual ETGs.
Middle and bottom rows in Figure~\ref{f8} show the large ETG  NGC~1407 observed by \citet{2012MNRAS.426.1475U} and \citetalias{2010AJ....139.1566F}.
\citetalias{2010AJ....139.1566F} observed CaT of 144 GCs in NGC~1407, and \citet{2012MNRAS.426.1475U} added more GCs and remeasured CaT for all GCs.  
The observation by \citetalias{2010AJ....139.1566F} shows that only the ${(B-I)_0}$ distribution has bimodality but the CaT distribution prefers skewed Gaussian distribution.
Our projection test based on 13~Gyr relations reproduces simultaneously both the $CaT$ and $(B-I)_0$ observations of \citetalias{2010AJ....139.1566F} using the Gaussian MDF of {${\rm \left<[Fe/H]\right>=-0.80}$} with $\sigma_{\rm [Fe/H]}=0.55$.
On the other hand, the GCs of \citet{2012MNRAS.426.1475U} show bimodality for both the $(g-i)_0$ and CaT distributions.
Our models for NGC~1407 show bimodal distributions both in $(g-i)_0$ and $CaT$.
However, the metal-poor bump (at ${CaT_{\rm F10}} \sim$ 4.8{\AA}) in the CaT distribution is not reproduced.
The parameters for these distributions analyzed by the GMM test are listed in Table~\ref{tab.2}.

In Figure~\ref{f9}, the simulated distributions for the GC systems in NGC~4278, NGC~4365, and NGC~3377 concur with the observations (see Table~\ref{tab.2}).
A small metal-rich bump (at ${CaT_{\rm F10}} \sim$ 7.5{\AA}) in the CaT distribution of NGC~4365 GCs seems related to the intermediate age GC populations reported by \citet{2003ApJ...585..767L, 2005A&A...443..413L} and \citet{2012MNRAS.420...37B}.
Our simulation is not successful in reproducing the distributions of NGC~3115 GC system.
Strongly separated two peaks in $(g-i)_0$ is hard to be reproduced under the assumption of a single Gaussian MDF.
As suggested by \citet{2012ApJ...759L..33B} and \citet{2014A&A...564L...3C}, this is because NGC~3115 has two metallicity groups of GCs confirmed by optical--to--NIR colors. 
{The detection of metal-poor stellar halo in NGC~3115 also provides a good explanation for two separate GC groups \citep{2015ApJ...800...13P}.}
Except for NGC~3115, our model reproduces the $(g-i)_0$ and CaT distributions {\it simultaneously} based on the $(g-i)_0$ CMR and CaT IMR that are nonlinear.

\begin{figure}
\includegraphics[angle=-90,scale=0.95]{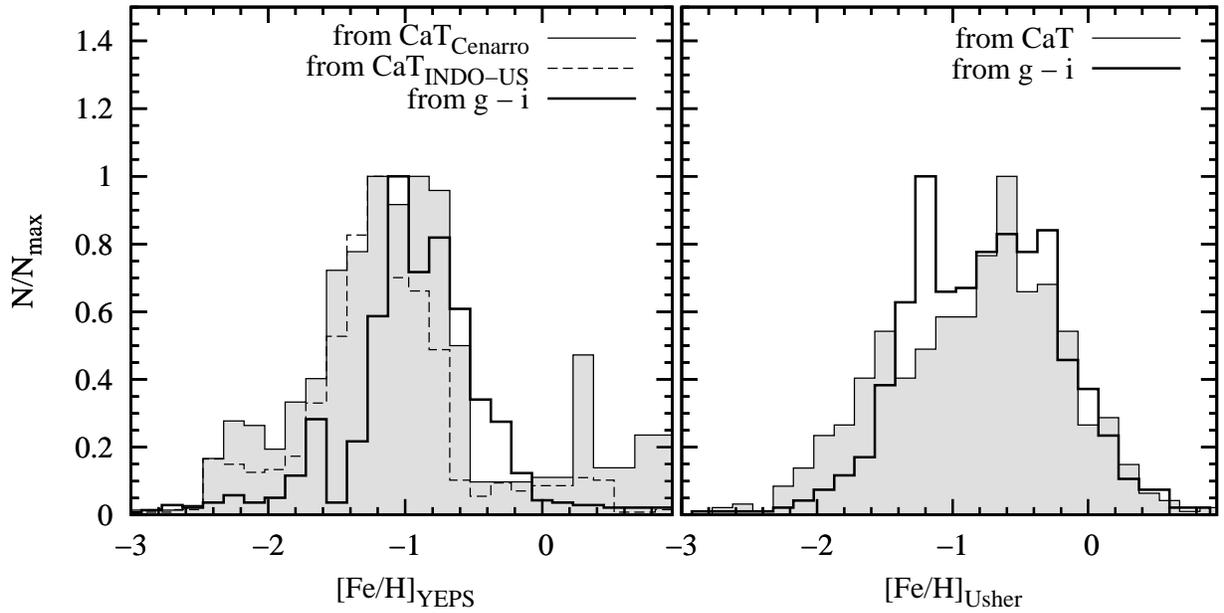}
\caption[]{
The comparison of derived GC MDFs based on different $CaT$--metallicity and $(g-i)_0$--metallicity relations. 
The left panel is the GC MDFs derived from our relations, and the right panel is from \citet{2012MNRAS.426.1475U}. 
Grey and empty histograms are derived GC MDFs from $CaT$ and $(g-i)_0$ color of 779 GCs, respectively. 
}
\label{f10}
\end{figure}

{Based on our CaT IMR and $(g-i)_0$ CMR, we have derived GC MDFs from $CaT$ and $(g-i)_0$ of 779 GCs.
Figure~\ref{f10} shows the result of the inverse projection and compares the derived GC MDFs with that presented in \citet{2012MNRAS.426.1475U}.
We exclude GCs of NGC~3115 in this simulation for reasons mentioned above.
For the conversion of ${\rm [Fe/H]_{Usher}}$ from $(g-i)_0$, we adopt Eq.~10 of \citet{2012MNRAS.426.1475U}. 
Both MDFs derived from our $CaT$ and $(g-i)_0$ models show similar skewed Gaussian shapes with a single peak around ${\rm [Fe/H]} \sim -1.0$. 
These shapes are consistent with the typical MDFs of field stars in ETGs \citepalias{2011ApJ...743..150Y}.  
Note that the MDFs from our $CaT$ model have small peaks where ${\rm [Fe/H]} \geq 0.2$.
These peaks are caused by the saturated features in CaT IMRs that make hard to determine the metallicity of metal-rich stellar populations.
The right panel, on the other hand, shows that the MDF derived from $CaT$ has two peaks at ${\rm [Fe/H]} = -1.41$ and $-0.43$ and the MDF from $(g-i)_0$ at ${\rm [Fe/H]} = -1.10$ and $-0.37$ based on the GMM test.
The two metallicity groups have different ratio between metal-poor and metal-rich GCs with 34.0 : 66.0\% and 47.1 : 52.9\%, respectively. 
}
 
The nonlinearity of the CaT IMR also have an important implication on the integrated CaT of various stellar systems.
As shown in our models, if cool stars are dominant in a given stellar system, the integrated CaT can not be the indicator of metallicity or even the calcium abundance.
We speculate that this is why calcium abundance derived from calcium II lines is unexpectedly low in various stellar systems.
For example, the observed CaT \citep{2002ApJ...579L..13S} and CaHK \citep{2011ApJ...729..148W} of ETGs suggest constant or even decreasing calcium abundances with increasing mass of galaxies. 
These results go against with the general trend of mass--metallicity relation for ETGs.
However, a larger fraction of cool metal-rich stars in the more massive ETGs could explain the observed CaT feature of ETGs.
We will fully discuss the issue in our forthcoming paper.

\acknowledgments

C.C. acknowledges support from the Research Fellow Program (NRF-2013R1A1A2006053) of the National Research Foundation of Korea.
Y.W.L. and S.J.Y. acknowledge support from the National Research Foundation of Korea to the Center for Galaxy Evolution Research (No. 2010-0027910).
S.J.Y. acknowledges support from Mid-career Researcher Program (No. 2015-008049) through the National Research Foundation (NRF) of Korea, and the Yonsei University Future-leading Research Initiative of 2014-2015.

\clearpage

\clearpage

\begin{figure}
\includegraphics[angle=-90,scale=0.9]{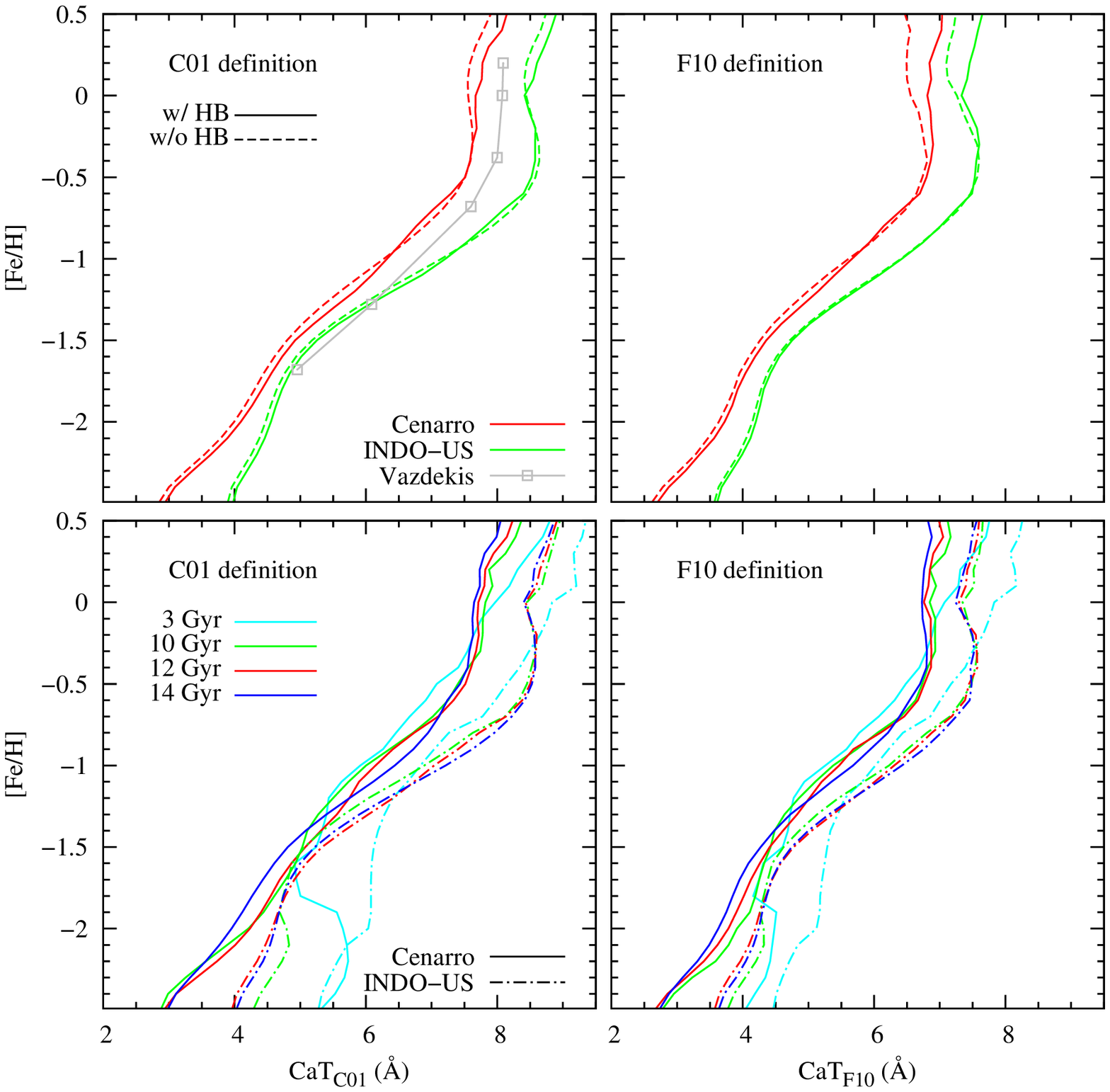}
\end{figure}
\begin{figure}
\caption[]{
The effect of HB stars and age of stellar populations on the simple stellar population model of CaT.
The CaT strengths in the left and right panels are measured based on the definitions of \citet{2001MNRAS.326..959C} and \citet{2010AJ....139.1566F}, respectively.
({\it Upper}) Red and green solid lines represent the CaT model with HB stars based on the Cenarro and INDO-US, respectively, at the age of 13~Gyr.
The dashed lines indicate the CaT model without HB stars.
{The 12.5893-Gyr model of \citet{2003MNRAS.340.1317V} is presented as a grey line with points for comparison.
Grey points of the \citet{2003MNRAS.340.1317V} model indicate ${\rm [Fe/H]}=-1.68$, $-1.28$, $-0.68$, $-0.38$, $0.00$, and $0.20$.}
({\it Lower}) Cyan, green, red, and blue lines are models for 3, 10, 12, and 14~Gyr, respectively.
Solid and dot dashed lines are models based on the Cenarro and INDO-US library, respectively.
The effects of HB stars and age of old stellar population on the CaT model are very small.
The CaT definition of \citetalias{2010AJ....139.1566F} strengthens the saturating shape of the $CaT$--metallicity relations in the metal-rich regime.

}
\label{f2}
\end{figure}

\clearpage

\clearpage

\clearpage

\clearpage

\clearpage

\clearpage

\begin{figure}
\includegraphics[angle=0,scale=0.8]{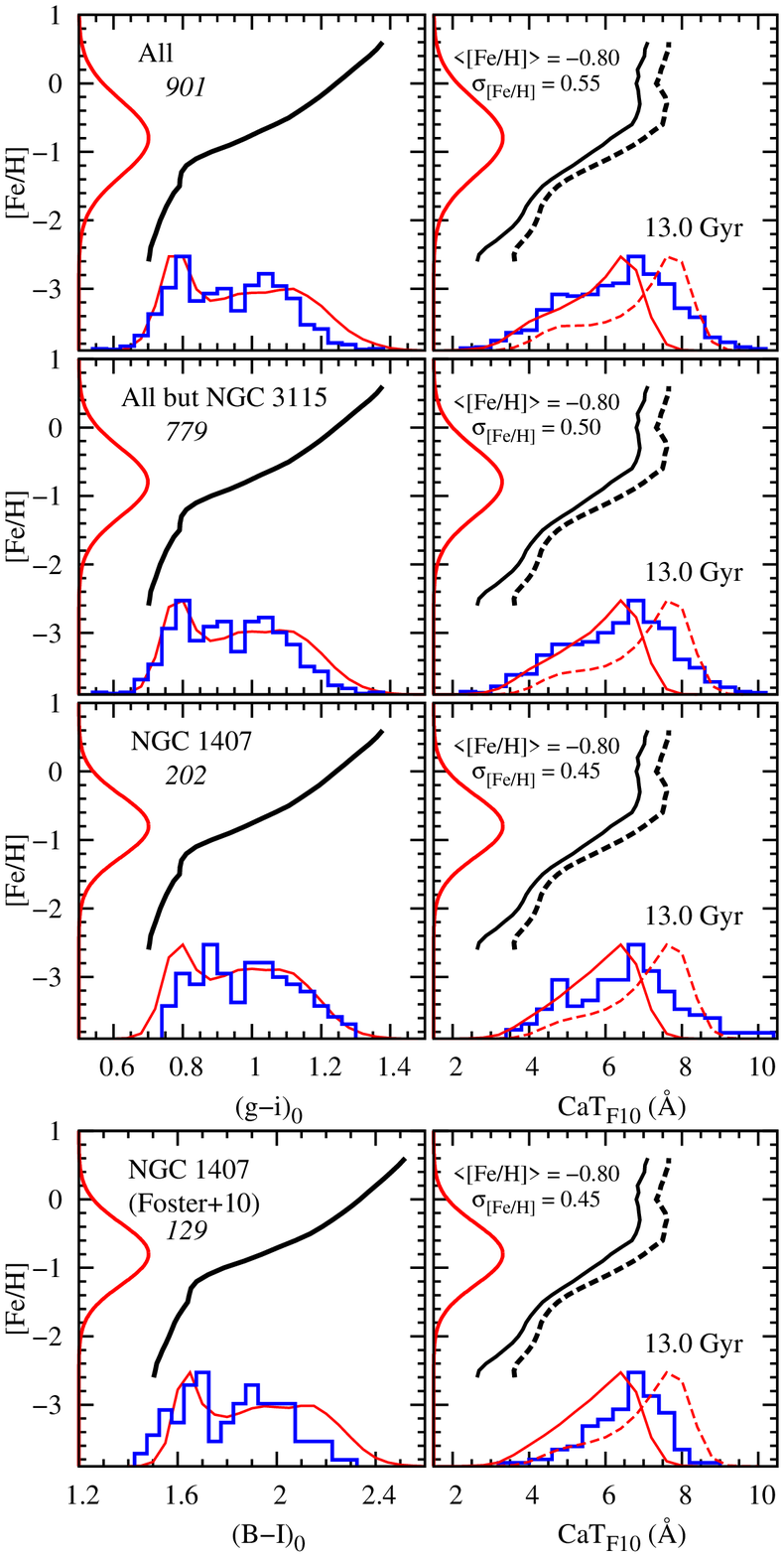}
\end{figure}

\begin{figure}
\caption[]{
The effect of the nonlinear nature of $CaT$--metallicity and {color--metallicity} relations on the projection of unimodal MDFs. 
The observed distribution (blue and open histograms) of GCs in ETGs are from \citet{2012MNRAS.426.1475U} and \citet{2010AJ....139.1566F}.
The black lines in each plot are the adopted CMRs and IMRs, and the black solid and dashed lines in the right panels are models based on Cenarro and INDO-US libraries, respectively.
The single Gaussian MDFs of $10^6$ model GCs are assumed, and the best-fit age, mean metallicity and standard deviation of ${\rm [Fe/H]}$ of each Gaussian MDF to reproduce the distributions of $(g-i)_0$, $(B-I)_0$, and $CaT_{\rm F10}$ simultaneously are indicated.
The red solid and dashed lines in the bottom of each plot show the result of the unimodal MDF projection to $(g-i)_0$, $(B-I)_0$, and $CaT_{\rm F10}$. 
Our models reproduce the overall shapes of both CaT and color distributions of GCs simultaneously.
}
\label{f8}
\end{figure}

\clearpage

\begin{figure}
\includegraphics[angle=0,scale=0.8]{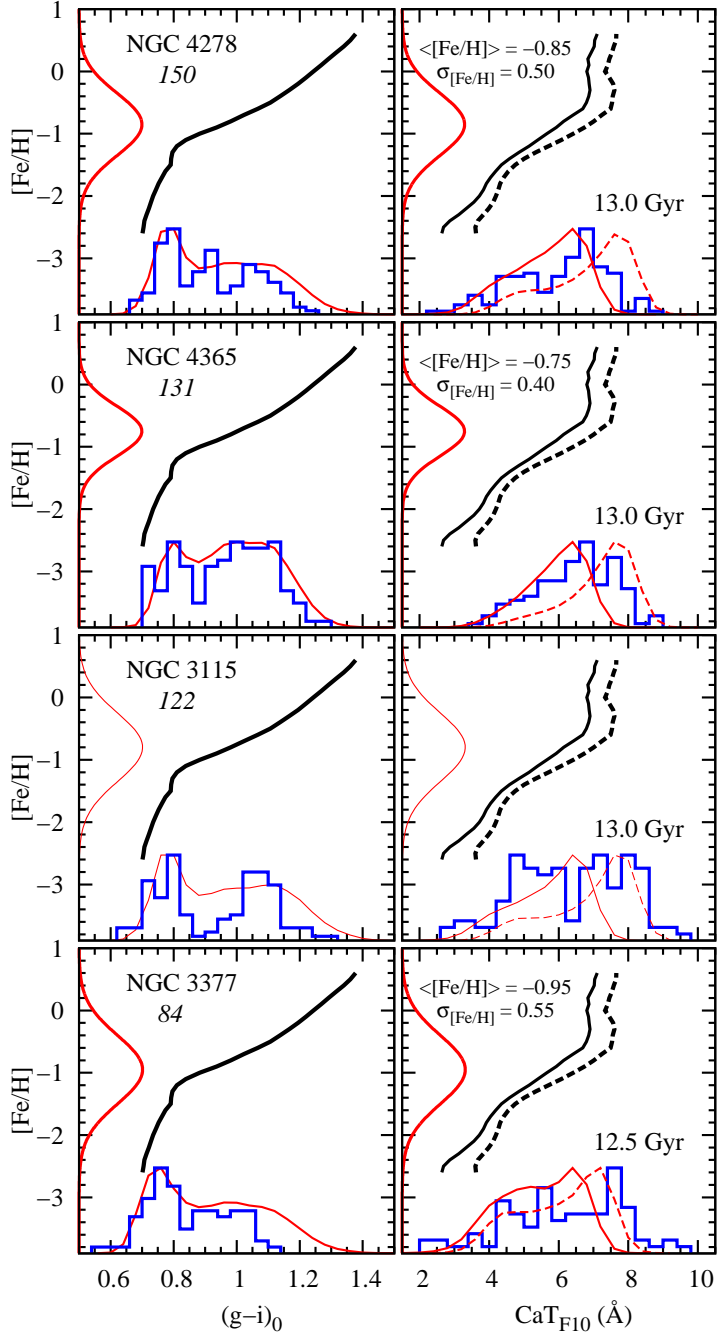}
\caption[]{
Same as Figure~\ref{f8} but for NGC~4278, NGC~4365, NGC~3115, and NGC~3377.
Our single Gaussian projection test is hard to reproduce $(g-i)_0$ and CaT distributions of GCs in NGC~3115 at the same time.
}
\label{f9}
\end{figure}

\clearpage

\clearpage

\begin{deluxetable}{r c cc c cc c cc c cc c cc}
\tablecaption{\label{tab.1}CaT of YEPS simple stellar population models for  [$\alpha$/Fe]=0.3 (based on empirical spectra of Cenarro and INDO-US library).}
\tabletypesize{\scriptsize}
\tablecolumns{16}
\tablewidth{0pt}
\tablehead{
\colhead{} & \colhead{} &
\multicolumn{2}{c}{$t=$ 3.0~(Gyr)} & \colhead{} &
\multicolumn{2}{c}{5.0} & \colhead{} &
\multicolumn{2}{c}{10.0} & \colhead{} &
\multicolumn{2}{c}{12.0} & \colhead{} &
\multicolumn{2}{c}{14.0}

\\
\cline{3-4} \cline{6-7} \cline{9-10} \cline{12-13} \cline{15-16}\\
\colhead{${\rm [Fe/H]}$} & \colhead{} &
\colhead{$CaT_{\rm C01}$} & \colhead{$CaT_{\rm F10}$} &\colhead{} &
\colhead{$CaT_{\rm C01}$} & \colhead{$CaT_{\rm F10}$} &\colhead{} &
\colhead{$CaT_{\rm C01}$} & \colhead{$CaT_{\rm F10}$} &\colhead{} & 
\colhead{$CaT_{\rm C01}$} & \colhead{$CaT_{\rm F10}$} &\colhead{} &
\colhead{$CaT_{\rm C01}$} & \colhead{$CaT_{\rm F10}$}
}

\startdata

\colhead{} & \colhead{} & \multicolumn{14}{c}{Cenarro library} \\
\hline

 -2.5 &&  5.30 &  4.03 &&  4.10 &  3.51 &&  2.87 &  2.76 &&  2.92 &  2.65 &&  2.97 &  2.72 \\
 -2.4 &&  5.52 &  4.18 &&  4.33 &  3.62 &&  2.98 &  2.95 &&  3.10 &  2.85 &&  3.10 &  2.87 \\
 -2.3 &&  5.67 &  4.34 &&  4.52 &  3.82 &&  3.25 &  3.23 &&  3.42 &  3.13 &&  3.32 &  3.08 \\
 -2.2 &&  5.72 &  4.41 &&  4.66 &  3.95 &&  3.56 &  3.58 &&  3.73 &  3.40 &&  3.55 &  3.30 \\
 -2.1 &&  5.70 &  4.44 &&  4.77 &  4.10 &&  3.88 &  3.78 &&  4.00 &  3.61 &&  3.76 &  3.48 \\
 -2.0 &&  5.64 &  4.48 &&  4.84 &  4.14 &&  4.21 &  3.91 &&  4.23 &  3.78 &&  3.95 &  3.62 \\
 -1.9 &&  5.55 &  4.51 &&  4.90 &  4.20 &&  4.43 &  4.10 &&  4.40 &  3.90 &&  4.11 &  3.74 \\
 -1.8 &&  5.00 &  4.15 &&  4.64 &  4.13 &&  4.60 &  4.19 &&  4.54 &  4.01 &&  4.26 &  3.84 \\
 -1.7 &&  4.94 &  4.24 &&  4.53 &  4.10 &&  4.77 &  4.24 &&  4.68 &  4.12 &&  4.42 &  3.95 \\
 -1.6 &&  4.91 &  4.31 &&  4.59 &  4.13 &&  4.92 &  4.32 &&  4.86 &  4.25 &&  4.60 &  4.08 \\
 -1.5 &&  5.24 &  4.60 &&  4.95 &  4.40 &&  5.03 &  4.42 &&  5.07 &  4.41 &&  4.81 &  4.27 \\
 -1.4 &&  5.33 &  4.68 &&  5.09 &  4.50 &&  5.10 &  4.48 &&  5.31 &  4.61 &&  5.07 &  4.47 \\
 -1.3 &&  5.39 &  4.71 &&  5.19 &  4.56 &&  5.27 &  4.63 &&  5.55 &  4.82 &&  5.39 &  4.72 \\
 -1.2 &&  5.43 &  4.77 &&  5.27 &  4.63 &&  5.49 &  4.85 &&  5.74 &  5.01 &&  5.75 &  5.04 \\
 -1.1 &&  5.62 &  4.93 &&  5.48 &  4.84 &&  5.73 &  5.10 &&  5.89 &  5.20 &&  6.10 &  5.34 \\
 -1.0 &&  5.91 &  5.25 &&  5.91 &  5.20 &&  6.00 &  5.37 &&  6.14 &  5.46 &&  6.43 &  5.67 \\
 -0.9 &&  6.26 &  5.57 &&  6.16 &  5.52 &&  6.35 &  5.73 &&  6.40 &  5.68 &&  6.72 &  5.94 \\
 -0.8 &&  6.45 &  5.77 &&  6.40 &  5.76 &&  6.71 &  6.05 &&  6.73 &  6.09 &&  6.94 &  6.21 \\
 -0.7 &&  6.66 &  6.06 &&  6.68 &  6.11 &&  7.01 &  6.40 &&  7.10 &  6.45 &&  7.11 &  6.38 \\
 -0.6 &&  6.91 &  6.31 &&  6.94 &  6.35 &&  7.25 &  6.62 &&  7.33 &  6.66 &&  7.24 &  6.53 \\
 -0.5 &&  7.07 &  6.48 &&  7.15 &  6.57 &&  7.40 &  6.73 &&  7.51 &  6.76 &&  7.43 &  6.69 \\
 -0.4 &&  7.40 &  6.68 &&  7.47 &  6.75 &&  7.55 &  6.81 &&  7.59 &  6.86 &&  7.55 &  6.79 \\
 -0.3 &&  7.54 &  6.79 &&  7.68 &  6.87 &&  7.74 &  6.92 &&  7.67 &  6.87 &&  7.57 &  6.80 \\
 -0.2 &&  7.63 &  6.88 &&  7.83 &  6.97 &&  7.77 &  6.92 &&  7.71 &  6.86 &&  7.63 &  6.78 \\
 -0.1 &&  7.76 &  6.93 &&  7.93 &  7.01 &&  7.78 &  6.93 &&  7.69 &  6.86 &&  7.62 &  6.73 \\
  0.0 &&  7.96 &  7.07 &&  8.10 &  7.07 &&  7.82 &  6.84 &&  7.71 &  6.75 &&  7.64 &  6.72 \\
  0.1 &&  8.18 &  7.28 &&  8.32 &  7.23 &&  7.92 &  6.94 &&  7.80 &  6.82 &&  7.73 &  6.74 \\
  0.2 &&  8.30 &  7.31 &&  8.32 &  7.25 &&  7.87 &  6.84 &&  7.81 &  6.84 &&  7.73 &  6.76 \\
  0.3 &&  8.49 &  7.51 &&  8.45 &  7.41 &&  8.12 &  7.07 &&  7.94 &  6.90 &&  7.80 &  6.81 \\
  0.4 &&  8.69 &  7.70 &&  8.56 &  7.49 &&  8.27 &  7.16 &&  8.14 &  7.04 &&  7.98 &  6.87 \\
  0.5 &&  8.79 &  7.75 &&  8.66 &  7.58 &&  8.36 &  7.12 &&  8.23 &  6.98 &&  8.05 &  6.81 \\

\hline
\colhead{} & \colhead{} & \multicolumn{14}{c}{INDO-US library} \\
\hline

 -2.5 &&  5.27 &  4.46 &&  4.93 &  4.23 &&  4.28 &  3.76 &&  3.95 &  3.56 &&  4.02 &  3.62 \\
 -2.4 &&  5.32 &  4.51 &&  5.04 &  4.33 &&  4.39 &  3.86 &&  4.03 &  3.64 &&  4.12 &  3.71 \\
 -2.3 &&  5.44 &  4.60 &&  5.24 &  4.48 &&  4.56 &  4.03 &&  4.18 &  3.79 &&  4.28 &  3.89 \\
 -2.2 &&  5.56 &  4.70 &&  5.42 &  4.63 &&  4.72 &  4.20 &&  4.34 &  3.96 &&  4.43 &  4.04 \\
 -2.1 &&  5.71 &  4.83 &&  5.53 &  4.75 &&  4.82 &  4.32 &&  4.46 &  4.09 &&  4.54 &  4.16 \\
 -2.0 &&  6.03 &  5.12 &&  5.76 &  4.96 &&  4.79 &  4.31 &&  4.57 &  4.19 &&  4.61 &  4.24 \\
 -1.9 &&  6.07 &  5.17 &&  5.70 &  4.94 &&  4.68 &  4.26 &&  4.65 &  4.26 &&  4.67 &  4.29 \\
 -1.8 &&  6.07 &  5.17 &&  5.59 &  4.89 &&  4.73 &  4.31 &&  4.77 &  4.35 &&  4.74 &  4.36 \\
 -1.7 &&  6.07 &  5.20 &&  5.49 &  4.85 &&  4.81 &  4.38 &&  4.90 &  4.44 &&  4.84 &  4.43 \\
 -1.6 &&  6.08 &  5.24 &&  5.42 &  4.80 &&  4.87 &  4.44 &&  5.09 &  4.57 &&  4.99 &  4.56 \\
 -1.5 &&  6.11 &  5.28 &&  5.41 &  4.80 &&  5.06 &  4.62 &&  5.33 &  4.77 &&  5.22 &  4.74 \\
 -1.4 &&  6.18 &  5.33 &&  5.49 &  4.88 &&  5.33 &  4.85 &&  5.66 &  5.04 &&  5.52 &  4.99 \\
 -1.3 &&  6.28 &  5.44 &&  5.64 &  5.05 &&  5.66 &  5.12 &&  6.03 &  5.36 &&  5.89 &  5.30 \\
 -1.2 &&  6.40 &  5.55 &&  5.87 &  5.26 &&  6.04 &  5.45 &&  6.41 &  5.69 &&  6.32 &  5.67 \\
 -1.1 &&  6.63 &  5.81 &&  6.22 &  5.60 &&  6.47 &  5.84 &&  6.73 &  6.00 &&  6.76 &  6.06 \\
 -1.0 &&  6.82 &  6.01 &&  6.57 &  5.92 &&  6.88 &  6.22 &&  7.03 &  6.30 &&  7.21 &  6.42 \\
 -0.9 &&  7.03 &  6.22 &&  6.93 &  6.23 &&  7.27 &  6.50 &&  7.40 &  6.61 &&  7.61 &  6.75 \\
 -0.8 &&  7.26 &  6.43 &&  7.30 &  6.52 &&  7.62 &  6.80 &&  7.72 &  6.87 &&  7.95 &  7.02 \\
 -0.7 &&  7.77 &  6.85 &&  7.88 &  6.92 &&  8.11 &  7.15 &&  8.13 &  7.18 &&  8.22 &  7.24 \\
 -0.6 &&  7.97 &  7.00 &&  8.14 &  7.10 &&  8.33 &  7.33 &&  8.37 &  7.38 &&  8.42 &  7.45 \\
 -0.5 &&  8.14 &  7.17 &&  8.31 &  7.33 &&  8.45 &  7.48 &&  8.50 &  7.44 &&  8.52 &  7.48 \\
 -0.4 &&  8.34 &  7.38 &&  8.45 &  7.46 &&  8.52 &  7.53 &&  8.56 &  7.57 &&  8.57 &  7.46 \\
 -0.3 &&  8.48 &  7.50 &&  8.59 &  7.61 &&  8.58 &  7.57 &&  8.58 &  7.56 &&  8.57 &  7.52 \\
 -0.2 &&  8.64 &  7.65 &&  8.61 &  7.62 &&  8.55 &  7.48 &&  8.60 &  7.55 &&  8.56 &  7.49 \\
 -0.1 &&  8.76 &  7.75 &&  8.61 &  7.62 &&  8.49 &  7.41 &&  8.49 &  7.37 &&  8.50 &  7.39 \\
  0.0 &&  8.83 &  7.82 &&  8.61 &  7.63 &&  8.45 &  7.34 &&  8.43 &  7.29 &&  8.40 &  7.23 \\
  0.1 &&  9.20 &  8.16 &&  9.04 &  7.95 &&  8.67 &  7.52 &&  8.59 &  7.39 &&  8.53 &  7.30 \\
  0.2 &&  9.17 &  8.14 &&  9.02 &  7.92 &&  8.74 &  7.51 &&  8.64 &  7.41 &&  8.56 &  7.37 \\
  0.3 &&  9.16 &  8.07 &&  9.05 &  7.92 &&  8.81 &  7.60 &&  8.74 &  7.51 &&  8.66 &  7.45 \\
  0.4 &&  9.29 &  8.20 &&  9.14 &  7.98 &&  8.88 &  7.64 &&  8.82 &  7.57 &&  8.78 &  7.48 \\
  0.5 &&  9.34 &  8.25 &&  9.20 &  8.03 &&  8.95 &  7.65 &&  8.90 &  7.59 &&  8.86 &  7.56 \\

\enddata
\end{deluxetable}

\clearpage

\begin{deluxetable}{lccccccrrr}
\tabletypesize{\scriptsize}
\tablewidth{0pt}
\tablecaption{\label{tab.2}GMM \citep{2010ApJ...718.1266M} analysis for distributions of Figures~\ref{f8} and \ref{f9}.}
\tablehead{
\colhead{All GCs} & \colhead{$\mu_1$} & \colhead{$\sigma_1$} & \colhead{$p_1$} & \colhead{$\mu_2$} & \colhead{$\sigma_2$} & \colhead{$p_2$} & \colhead{$P(\chi^2)$} & \colhead{$P(DD)$} & \colhead{$P(kurt)$}}
\startdata
$CaT_{\rm Usher \: et \:al. \: (2012)}$   &  5.087 & 0.9359 & 0.3428 & 7.239 & 0.9749 & 0.6572 & $<0.001$ & 0.184 & 0.001  \\ 
$CaT_{\rm Model \: (Cenarro)}$            &  5.224 & 0.9263 & 0.5325 & 6.704 & 0.4628 & 0.4675 & $<0.010$ & 0.010 & $<0.010$ \\ 
$CaT_{\rm Model \: (INDO-US)}$            &  5.994 & 1.0240 & 0.4103 & 7.902 & 0.4990 & 0.5897 & $<0.010$ & 0.040 & $<0.010$ \\ 
$(g-i)_{\rm Usher \: et \:al. \: (2012)}$ &  0.812 & 0.0688 & 0.3967 & 1.051 & 0.1045 & 0.6033 & $<0.001$ & 0.115 & $<0.001$  \\ 
$(g-i)_{\rm Model}$                       &  0.802 & 0.0461 & 0.3110 & 1.071 & 0.1391 & 0.6890 & $<0.010$ & 0.010 & $<0.010$ \\ 
\hline
\colhead{All but NGC~3115} & \colhead{$\mu_1$} & \colhead{$\sigma_1$} & \colhead{$p_1$} & \colhead{$\mu_2$} & \colhead{$\sigma_2$} & \colhead{$p_2$} & \colhead{$P(\chi^2)$} & \colhead{$P(DD)$} & \colhead{$P(kurt)$}\\ 
\hline
$CaT_{\rm Usher \: et \:al. \: (2012)}$   &  4.868 & 0.8437 & 0.2713 & 7.090 & 1.0084 & 0.7287 & $<0.001$ & 0.175 & 0.009  \\ 
$CaT_{\rm Model \: (Cenarro)}$            &  5.329 & 0.9002 & 0.5332 & 6.689 & 0.4650 & 0.4668 & $<0.010$ & 0.030 & $<0.010$ \\ 
$CaT_{\rm Model \: (INDO-US)}$            &  6.076 & 1.0035 & 0.3807 & 7.877 & 0.5202 & 0.6193 & $<0.010$ & 0.210 & 0.030 \\ 
$(g-i)_{\rm Usher \: et \:al. \: (2012)}$ &  0.817 & 0.0700 & 0.3701 & 1.041 & 0.1115 & 0.6299 & $<0.001$ & 0.145 & $<0.001$  \\ 
$(g-i)_{\rm Model}$                       &  0.806 & 0.0461 & 0.3038 & 1.064 & 0.1299 & 0.6962 & $<0.010$ & $<0.010$ & $<0.010$ \\ 
\hline
\colhead{NGC~1407} & \colhead{$\mu_1$} & \colhead{$\sigma_1$} & \colhead{$p_1$} & \colhead{$\mu_2$} & \colhead{$\sigma_2$} & \colhead{$p_2$} & \colhead{$P(\chi^2)$} & \colhead{$P(DD)$} & \colhead{$P(kurt)$}\\ 
\hline
$CaT_{\rm Usher \: et \:al. \: (2012)}$   &  4.848 & 0.4991 & 0.1575 & 6.995 & 1.1276 & 0.8425 & 0.054 & 0.279 & 0.119  \\ 
$CaT_{\rm Foster \: et \:al. \: (2010)}$  &  5.959 & 0.7592 & 0.3757 & 7.235 & 0.5688 & 0.6243 & 0.050 & 0.600 & 0.556  \\ 
$CaT_{\rm Model \: (Cenarro)}$            &  5.360 & 0.8525 & 0.5231 & 6.644 & 0.4737 & 0.4769 & $<0.010$ & 0.010 & $<0.010$ \\ 
$CaT_{\rm Model \: (INDO-US)}$            &  6.216 & 0.9859 & 0.3947 & 7.869 & 0.5116 & 0.6053 & $<0.010$ & 0.160 & 0.490 \\ 
$(g-i)_{\rm Usher \: et \:al. \: (2012)}$ &  0.864 & 0.0540 & 0.3329 & 1.076 & 0.1090 & 0.6671 & $<0.001$ & 0.297 & $<0.001$  \\ 
$(g-i)_{\rm Model}$                       &  0.808 & 0.0429 & 0.2723 & 1.047 & 0.1255 & 0.7277 & $<0.010$ & 0.010 & $<0.010$ \\ 
${(B-I)}_{\rm Foster \: et \:al. \: (2010)}$&  1.650 & 0.0812 & 0.3761 & 1.973 & 0.1441 & 0.6239 & 0.001 & 0.248 & $<0.001$  \\ 
${(B-I)}_{\rm Model}$                       &  1.669 & 0.0546 & 0.2625 & 2.038 & 0.1892 & 0.7375 & $<0.010$ & $<0.010$ & $<0.010$ \\ 
\hline

\colhead{NGC~4278} & \colhead{$\mu_1$} & \colhead{$\sigma_1$} & \colhead{$p_1$} & \colhead{$\mu_2$} & \colhead{$\sigma_2$} & \colhead{$p_2$} & \colhead{$P(\chi^2)$} & \colhead{$P(DD)$} & \colhead{$P(kurt)$}\\ 
\hline
$CaT_{\rm Usher \: et \:al. \: (2012)}$   &  4.939 & 0.9225 & 0.4074 & 7.097 & 0.6919 & 0.5926 & $<0.001$ & 0.266 & 0.130 \\ 
$CaT_{\rm Model \: (Cenarro)}$            &  5.223 & 0.9005 & 0.5763 & 6.688 & 0.4535 & 0.4237 & $<0.010$ & $<0.010$ & $<0.010$  \\ 
$CaT_{\rm Model \: (INDO-US)}$            &  5.989 & 0.9944 & 0.4409 & 7.883 & 0.5070 & 0.5591 & $<0.010$ & 0.020 & $<0.010$  \\ 
$(g-i)_{\rm Usher \: et \:al. \: (2012)}$ &  0.796 & 0.0497 & 0.3766 & 1.014 & 0.1108 & 0.6234 & $<0.001$ & 0.299 & $<0.001$  \\ 
$(g-i)_{\rm Model}$                       &  0.824 & 0.0495 & 0.4372 & 1.080 & 0.1088 & 0.5628 & $<0.010$ & 0.010 & $<0.010$  \\ 
\hline
\colhead{NGC~4365} & \colhead{$\mu_1$} & \colhead{$\sigma_1$} & \colhead{$p_1$} & \colhead{$\mu_2$} & \colhead{$\sigma_2$} & \colhead{$p_2$} & \colhead{$P(\chi^2)$} & \colhead{$P(DD)$} & \colhead{$P(kurt)$}\\ 
\hline
$CaT_{\rm Usher \: et \:al. \: (2012)}$   &  4.945 & 0.5368 & 0.1786 & 7.026 & 0.8675 & 0.8214 & 0.046 & 0.201 & 0.069 \\ 
$CaT_{\rm Model \: (Cenarro)}$            &  5.593 & 0.8056 & 0.5228 & 6.691 & 0.4405 & 0.4772 & $<0.010$ & 0.010 & 0.040  \\ 
$CaT_{\rm Model \: (INDO-US)}$            &  6.538 & 0.9549 & 0.3763 & 7.913 & 0.4874 & 0.6237 & $<0.010$ & 0.300 & 1.000  \\ 
$(g-i)_{\rm Usher \: et \:al. \: (2012)}$ &  0.796 & 0.0440 & 0.2755 & 1.048 & 0.0972 & 0.7245 & $<0.001$ & 0.122 & $<0.001$  \\ 
$(g-i)_{\rm Model}$                       &  0.820 & 0.0446 & 0.2322 & 1.055 & 0.1162 & 0.7678 & $<0.010$ & $<0.010$ & $<0.010$  \\ 

\hline
\colhead{NGC~3115} & \colhead{$\mu_1$} & \colhead{$\sigma_1$} & \colhead{$p_1$} & \colhead{$\mu_2$} & \colhead{$\sigma_2$} & \colhead{$p_2$} & \colhead{$P(\chi^2)$} & \colhead{$P(DD)$} & \colhead{$P(kurt)$}\\ 
\hline
$CaT_{\rm Usher \: et \:al. \: (2012)}$   & 5.551 & 1.0878 & 0.5711 & 7.983 & 0.6911 & 0.4289 & 0.002 & 0.262 & 0.018 \\ 
$(g-i)_{\rm Usher \: et \:al. \: (2012)}$ & 0.785 & 0.0576 & 0.4727 & 1.087 & 0.0694 & 0.5273 & $<0.001$ & 0.024 & $<0.001$  \\ 
\hline
\colhead{NGC~3377} & \colhead{$\mu_1$} & \colhead{$\sigma_1$} & \colhead{$p_1$} & \colhead{$\mu_2$} & \colhead{$\sigma_2$} & \colhead{$p_2$} & \colhead{$P(\chi^2)$} & \colhead{$P(DD)$} & \colhead{$P(kurt)$}\\ 
\hline
$CaT_{\rm Usher \: et \:al. \: (2012)}$   &  5.341 & 1.1644 & 0.5560 & 7.794 & 0.7629 & 0.4440  & 0.085 & 0.432 & 0.098 \\ 
$CaT_{\rm Model \: (Cenarro)}$            &  5.056 & 0.9288 & 0.6596 & 6.690 & 0.4434 & 0.3404  & $<0.010$ & 0.010 & $<0.010$  \\ 
$CaT_{\rm Model \: (INDO-US)}$            &  5.302 & 0.8913 & 0.5060 & 7.235 & 0.5057 & 0.4940  & $<0.010$ & $<0.010$ & $<0.010$  \\ 
$(g-i)_{\rm Usher \: et \:al. \: (2012)}$ &  0.788 & 0.0793 & 0.6642 & 1.013 & 0.0556 & 0.3358  & 0.006 & 0.154 & 0.007  \\ 
$(g-i)_{\rm Model}$                       &  0.772 & 0.0523 & 0.3615 & 1.031 & 0.1384 & 0.6385  & $<0.010$ & $<0.010$ & $<0.010$  \\ 
\enddata
\tablecomments{The mean, standard deviation, and portion of two groups analyzed by the GMM test are expressed as $\mu_{1,2}$, $\sigma_{1,2}$, and $p_{1,2}$, respectively.
The statistical significance of bimodality are expressed as $P$-values in the last three columns.
If $P$ is less than 0.05, the GMM test strongly suggests the bimodality of a distribution.
}
\end{deluxetable}

\clearpage


\begin{thebibliography}{}

\bibitem[Arnold et al.(2011)]{2011ApJ...736L..26A} Arnold, J.~A., Romanowsky, A.~J., Brodie, J.~P., et al.\ 2011, \apjl, 736, L26
\bibitem[Armandroff \& Zinn(1988)]{1988AJ.....96...92A} Armandroff, T.~E., \& Zinn, R.\ 1988, \aj, 96, 92
\bibitem[Battaglia et al.(2008)]{2008MNRAS.383..183B} Battaglia, G., Irwin, M., Tolstoy, E., et al.\ 2008, \mnras, 383, 183 
\bibitem[Beasley et al.(2008)]{2008MNRAS.386.1443B} Beasley, M.~A., Bridges, T., Peng, E., et al.\ 2008, \mnras, 386, 1443
\bibitem[Bica \& Alloin(1987)]{1987A&A...186...49B} Bica, E., \& Alloin, D.\ 1987, \aap, 186, 49
\bibitem[Blom et al.(2012)]{2012MNRAS.420...37B} Blom, C., Spitler, L.~R., \& Forbes, D.~A.\ 2012, \mnras, 420, 37
\bibitem[Brodie \& Strader(2006)]{2006ARA&A..44..193B} Brodie, J.~P., \& Strader, J.\ 2006, \araa, 44, 193
\bibitem[Brodie et al.(2012)]{2012ApJ...759L..33B} Brodie, J.~P., Usher, C., Conroy, C., et al.\ 2012, \apjl, 759, L33 
\bibitem[Cantiello et al.(2014)]{2014A&A...564L...3C} Cantiello, M., Blakeslee, J.~P., Raimondo, G., et al.\ 2014, \aap, 564, LL3
\bibitem[Carrera(2012)]{2012A&A...544A.109C} Carrera, R.\ 2012, \aap, 544, A109 
\bibitem[Cenarro et al.(2007)]{2007AJ....134..391C} Cenarro, A.~J., Beasley, M.~A., Strader, J., Brodie, J.~P., \& Forbes, D.~A.\ 2007, \aj, 134, 391
\bibitem[Cenarro et al.(2001a)]{2001MNRAS.326..959C} Cenarro, A.~J., Cardiel, N., Gorgas, J., et al.\ 2001, \mnras, 326, 959 
\bibitem[Cenarro et al.(2001b)]{2001MNRAS.326..981C} Cenarro, A.~J., Gorgas, J., Cardiel, N., et al.\ 2001, \mnras, 326, 981 
\bibitem[Cenarro et al.(2002)]{2002MNRAS.329..863C} Cenarro, A.~J., Gorgas, J., Cardiel, N., Vazdekis, A., \& Peletier, R.~F.\ 2002, \mnras, 329, 863 
\bibitem[Cenarro et al.(2007)]{2007MNRAS.374..664C} Cenarro, A.~J., Peletier, R.~F., S{\'a}nchez-Bl{\'a}zquez, P., et al.\ 2007, \mnras, 374, 664 
\bibitem[Chies-Santos et al.(2012)]{2012MNRAS.427.2349C} Chies-Santos, A.~L., Larsen, S.~S., \& Kissler-Patig, M.\ 2012, \mnras, 427, 2349
\bibitem[Chung et al.(2013b)]{2013ApJ...769L...3C} Chung, C., Lee, S.-Y., Yoon, S.-J., \& Lee, Y.-W.\ 2013, \apjl, 769, L3 
\bibitem[Chung et al.(2013a)]{2013ApJS..204....3C} Chung, C., Yoon, S.-J., Lee, S.-Y., \& Lee, Y.-W.\ 2013, \apjs, 204, 3
\bibitem[Delisle \& Hardy(1992)]{1992AJ....103..711D} Delisle, S., \& Hardy, E.\ 1992, \aj, 103, 711
\bibitem[Diaz et al.(1989)]{1989MNRAS.239..325D} Diaz, A.~I., Terlevich, E., \& Terlevich, R.\ 1989, \mnras, 239, 325
\bibitem[Foster et al.(2010)]{2010AJ....139.1566F} Foster, C., Forbes, D.~A., Proctor, R.~N., et al.\ 2010, \aj, 139, 1566 
\bibitem[Foster et al.(2011)]{2011MNRAS.415.3393F} Foster, C., Spitler, L.~R., Romanowsky, A.~J., et al.\ 2011, \mnras, 415, 3393
\bibitem[Harris(1991)]{1991ARA&A..29..543H} Harris, W.~E.\ 1991, \araa, 29, 543
\bibitem[Jennings et al.(2014)]{2014AJ....148...32J} Jennings, Z.~G., Strader, J., Romanowsky, A.~J., et al.\ 2014, \aj, 148, 32
\bibitem[Jones et al.(1984)]{1984ApJ...283..457J} Jones, J.~E., Alloin, D.~M., \& Jones, B.~J.~T.\ 1984, \apj, 283, 457
\bibitem[Kim et al.(2013, Paper V)]{2013ApJ...768..138K} Kim, S., Yoon, S.-J., Chung, C., et al.\ 2013, \apj, 768, 138
\bibitem[Kim et al.(2002)]{2002ApJS..143..499K} Kim, Y.-C., Demarque, P., Yi, S.~K., \& Alexander, D.~R.\ 2002, \apjs, 143, 499 
\bibitem[Larsen et al.(2003)]{2003ApJ...585..767L} Larsen, S.~S., Brodie, J.~P., Beasley, M.~A., et al.\ 2003, \apj, 585, 767
\bibitem[Larsen et al.(2005)]{2005A&A...443..413L} Larsen, S.~S., Brodie, J.~P., \& Strader, J.\ 2005, \aap, 443, 413 
\bibitem[Lee \& Worthey(2005)]{2005ApJS..160..176L} Lee, H.-c., \& Worthey, G.\ 2005, \apjs, 160, 176
\bibitem[Muratov \& Gnedin(2010)]{2010ApJ...718.1266M} Muratov, A.~L., \& Gnedin, O.~Y.\ 2010, \apj, 718, 1266
\bibitem[Park et al.(2012a)]{2012ApJ...759..116P} Park, H.~S., Lee, M.~G., Hwang, H.~S., et al.\ 2012, \apj, 759, 116
\bibitem[Park et al.(2012b)]{2012ApJ...757..184P} Park, H.~S., Lee, M.~G., \& Hwang, H.~S.\ 2012, \apj, 757, 184
\bibitem[Peng et al.(2004)]{2004ApJ...602..705P} Peng, E.~W., Ford, H.~C., \& Freeman, K.~C.\ 2004, \apj, 602, 705
\bibitem[Peacock et al.(2015)]{2015ApJ...800...13P} Peacock, M.~B., Strader, J., Romanowsky, A.~J., \& Brodie, J.~P.\ 2015, \apj, 800, 13
\bibitem[Rutledge et al.(1997)]{1997PASP..109..883R} Rutledge, G.~A., Hesser, J.~E., Stetson, P.~B., et al.\ 1997, \pasp, 109, 883
\bibitem[Saglia et al.(2002)]{2002ApJ...579L..13S} Saglia, R.~P., Maraston, C., Thomas, D., Bender, R., \& Colless, M.\ 2002, \apjl, 579, L13 
\bibitem[S{\'a}nchez-Bl{\'a}zquez et al.(2006)]{2006MNRAS.371..703S} S{\'a}nchez-Bl{\'a}zquez, P., Peletier, R.~F., Jim{\'e}nez-Vicente, J., et al.\ 2006, \mnras, 371, 703 
\bibitem[Sansom et al.(2013)]{2013MNRAS.435..952S} Sansom, A.~E., de Castro Milone, A., Vazdekis, A., \& S{\'a}nchez-Bl{\'a}zquez, P.\ 2013, \mnras, 435, 952
\bibitem[Strader et al.(2005)]{2005AJ....130.1315S} Strader, J., Brodie, J.~P., Cenarro, A.~J., Beasley, M.~A., \& Forbes, D.~A.\ 2005, \aj, 130, 1315
\bibitem[Thomas et al.(2003)]{2003MNRAS.339..897T} Thomas, D., Maraston, C., \& Bender, R.\ 2003, \mnras, 339, 897
\bibitem[Usher et al.(2012)]{2012MNRAS.426.1475U} Usher, C., Forbes, D.~A., Brodie, J.~P., et al.\ 2012, \mnras, 426, 1475 
\bibitem[Usher et al.(2015)]{2015MNRAS.446..369U} Usher, C., Forbes, D.~A., Brodie, J.~P., et al.\ 2015, \mnras, 446, 369
\bibitem[Usher et al.(2013)]{2013MNRAS.436.1172U} Usher, C., Forbes, D.~A., Spitler, L.~R., et al.\ 2013, \mnras, 436, 1172 
\bibitem[Valdes et al.(2004)]{2004ApJS..152..251V} Valdes, F., Gupta, R., Rose, J.~A., Singh, H.~P., \& Bell, D.~J.\ 2004, \apjs, 152, 251 
\bibitem[Vasquez et al.(2015)]{2015arXiv150700425V} Vasquez, S., Zoccali, M., Hill, V., et al.\ 2015, arXiv:1507.00425
\bibitem[Vazdekis et al.(2003)]{2003MNRAS.340.1317V} Vazdekis, A., Cenarro, A.~J., Gorgas, J., Cardiel, N., \& Peletier, R.~F.\ 2003, \mnras, 340, 1317
\bibitem[Vazdekis et al.(2015)]{2015MNRAS.449.1177V} Vazdekis, A., Coelho, P., Cassisi, S., et al.\ 2015, \mnras, 449, 1177
\bibitem[Vazdekis et al.(2012)]{2012MNRAS.424..157V} Vazdekis, A., Ricciardelli, E., Cenarro, A.~J., et al.\ 2012, \mnras, 424, 157 
\bibitem[Warren \& Cole(2009)]{2009MNRAS.393..272W} Warren, S.~R., \& Cole, A.~A.\ 2009, \mnras, 393, 272
\bibitem[West et al.(2004)]{2004Natur.427...31W} West, M.~J., C{\^o}t{\'e}, P., Marzke, R.~O., \& Jord{\'a}n, A.\ 2004, \nat, 427, 31
\bibitem[Westera et al.(2002)]{2002A&A...381..524W} Westera, P., Lejeune, T., Buser, R., Cuisinier, F., \& Bruzual, G.\ 2002, \aap, 381, 524
\bibitem[Woodley et al.(2010)]{2010ApJ...708.1335W} Woodley, K.~A., Harris, W.~E., Puzia, T.~H., et al.\ 2010, \apj, 708, 1335
\bibitem[Worthey et al.(2011)]{2011ApJ...729..148W} Worthey, G., Ingermann, B.~A., \& Serven, J.\ 2011, \apj, 729, 148 
\bibitem[Yoon et al.(2011b, Paper III)]{2011ApJ...743..150Y} Yoon, S.-J., Lee, S.-Y., Blakeslee, J.~P., et al.\ 2011, \apj, 743, 150
\bibitem[Yoon et al.(2011a, Paper II)]{2011ApJ...743..149Y} Yoon, S.-J., Sohn, S.~T., Lee, S.-Y., et al.\ 2011, \apj, 743, 149
\bibitem[Yoon et al.(2013, Paper IV)]{2013ApJ...768..137Y} Yoon, S.-J., Sohn, S.~T., Kim, H.-S., et al.\ 2013, \apj, 768, 137
\bibitem[Yoon et al.(2006, Paper I)]{2006Sci...311.1129Y} Yoon, S.-J., Yi, S.~K., \& Lee, Y.-W.\ 2006, Science, 311, 1129
\bibitem[Zhou(1991)]{1991A&A...248..367Z} Zhou, X.\ 1991, \aap, 248, 367

\end{thebibliography}
\end{document}